\RequirePackage{fix-cm}
\documentclass[twocolumn,epjc3,showpacs,preprintnumbers,amsmath,amssymb]{svjour3}
\smartqed  
\RequirePackage{graphicx}
\RequirePackage[numbers,sort&compress]{natbib}
\usepackage{graphicx,amsmath,amsfonts,latexsym,amssymb,graphics,epsfig,subfigure,color,makeidx,hyperref}
\usepackage{multirow}
\usepackage{mathtools, cuted}

\journalname{Eur. Phys. J. C}

\newcommand\beq{\begin{equation}}
\newcommand\eeq{\end{equation}}
\newcommand\beqn{\begin{eqnarray}}
\newcommand\eeqn{\end{eqnarray}}
\newcommand\bal{\begin{align}}
\newcommand\eal{\end{align}}

\begin{document}

\allowdisplaybreaks

\title{Equivalence of solutions between the four-dimensional novel and regularized EGB theories in a cylindrically symmetric spacetime
}

\author{Zi-Chao Lin\thanksref{addr1,e1} \and Ke Yang\thanksref{addr2,e2} \and Shao-Wen Wei\thanksref{addr1,e3} \and Yong-Qiang Wang\thanksref{addr1,e4} \and Yu-Xiao Liu\thanksref{addr1,addr3,e5}
}                    

\thankstext{e1}{linzch18@lzu.edu.cn}
\thankstext{e2}{keyang@swu.edu.cn}
\thankstext{e3}{weishw@lzu.edu.cn}
\thankstext{e4}{yqwang@lzu.edu.cn}
\thankstext{e5}{liuyx@lzu.edu.cn, corresponding author}

\institute{
Institute of Theoretical Physics $\&$ Research Center of Gravitation, Lanzhou University, Lanzhou 730000, China \label{addr1}
\and
School of Physical Science and Technology, Southwest University, Chongqing 400715, China\label{addr2}
\and
Key Laboratory for Magnetism and Magnetic of the Ministry of Education, Lanzhou University, Lanzhou 730000, China\label{addr3}
 }

\maketitle

\abstract{
 Recently, a novel four-dimensional Einstein-Gauss-Bonnet (EGB) theory was presented to bypass the Lovelock's theorem and to give nontrivial effects on the four-dimensional local gravity. The main mechanism is to introduce a redefinition $\alpha\rightarrow\alpha/(D-4)$ and to take the limit $D\rightarrow4$. However, this theory does not have standard four-dimensional field equations. Some regularization procedures are then proposed to address this problem [arXiv:2003.11552, arXiv:2003.12771, arXiv:2004.08362, arXiv:2004.09472, arXiv:2004.10716]. The resultant regularized four-dimensional EGB theory has the same on-shell action as the original theory. Thus it is expected that the novel four-dimensional EGB theory is equivalent to its regularized version. However, the equivalence of these two theories is symmetry-dependent. In this paper, we test the equivalence in a cylindrically symmetric spacetime. The well-defined field equations of the two theories are obtained, with which our follow-up analysis shows that they are equivalent in such spacetime. Cylindrical cosmic strings are then considered as specific examples of the metric. Three sets of solutions are obtained and the corresponding string mass densities are evaluated. The results reveal how the Gauss-Bonnet term in four dimensions contributes to the string geometry in the new theory.
}



\maketitle

\section{Introduction}

The Einstein-Gauss-Bonnet (EGB) theory is a particular generalization of general relativity (GR), where a Gauss-Bonnet (GB) term constituted of higher curvature terms is introduced to GR. One main feature of such theory is that the GB term becomes a topological invariant when the spacetime dimension equals to four. Therefore, no contribution from the GB term could be found in the field equations. However, recent researches on the EGB theory showed that there are nontrivial effects from the GB term by introducing the redefinition $\alpha\rightarrow\alpha/(D-4)$ and taking the limit $D\rightarrow4$ ~\cite{Tomozawa2011,Cognola2013,Glavan1}. The resultant theory is now dubbed as the novel four-dimensional (4D) EGB theory. Some related investigations can be found in Refs.~\cite{Konoplya1,Fernandes1,Wei1,Zhang1,Lu1,Heydari-Fard1,Hennigar1,Jusufi2,Mahapatra1,Yang1,Yang2,Guo1,Hennigar2,Hennigar3,Odintsov1,Odintsov2,Oikonomou1,Easson1}. Indeed, when one investigates nontrivial effects in the novel 4D EGB gravity, the above approach requires one to start from a higher-dimensional version and to take the $D\rightarrow4$ limit. However, the authors in Ref.~\cite{Gurses1} pointed out that the 4D EGB gravity does not generally admit an intrinsically four-dimensional definition since, for a generic metic, the Bianchi identity is not satisfied when $D=4$~\footnote{An equivalent statement could be that, for a generic metric, not all parts of the GB tensor contain the $(D-4)$ factor~\cite{Gurses1}. The other argument of the authors is that, in the torsion free case, the GB term does not carry the vielbein when $D=4$. Thus in the vielbein formulation, the Einstein's equations are recovered.}. To prevent the discontinuity of the Lanczos-Bach tensor and to satisfy the Bianchi identity in the redefinition $\alpha\rightarrow\alpha/(D-4)$ and limit $D\rightarrow4$, only the solutions that make the Bianchi identity on-shell are allowed. Thus generally speaking, the theory only makes sense on some selected highly-symmetric spacetimes, such as the static spherically symmetric spacetime and FLRW spacetime, but it is ill-defined in four dimensions with the dimensional regularization procedure, such as the homogeneous but anisotropic Bianchi type I spacetime, whose evolution was found to be dependent on the choice of the symmetry between the extra nonphysical dimensions and the physical spatial dimensions~\cite{Tian2020}.

To address this issue, some regularized 4D EGB theories were constructed~\cite{Fernandes2,Lu1,Hennigar1,Kobayashi1,Bonifacio1}. The regularization procedure is not unique~\footnote{Here and after, the referred regularization procedures only involve the ones that introduce an additional degree of freedom while preserve the diffeomorphism invariance. For the $D\rightarrow4$ EGB theory that breaks the diffeomorphism invariance, see Ref.~\cite{Aoki1}.}. One mechanism is to introduce an extra GB term constructed by a conformal metric~\cite{Fernandes2,Hennigar1}. The resultant gravity has an extra scalar degree of freedom which helps us to obtain the full field equations in four dimensions. Another mechanism to obtain a well-defined 4D version of the EGB theory was introduced in Refs.~\cite{Lu1,Kobayashi1,Bonifacio1}. Therein, the Kaluza-Klein dimensional reduction of the $D$-dimensional EGB theory on a $(D-4)$-dimensional maximally symmetric space was used. Indeed, both these mechanisms could give the same regularized 4D EGB theory~\cite{Fernandes2,Easson1} which belongs to the Horndeski theory~\cite{Lu1}. Obviously, the regularized 4D EGB theory is dramatically different from the original one. And it will never recover the novel 4D EGB theory unless the extra scalar degree of freedom is ``missing''~\cite{Easson1}. Nevertheless, there is an interesting feature between these two theories.

As was pointed out in Ref.~\cite{Fernandes2} that the extra GB term constructed by the conformal metric vanishes on-shell. Thus the on-shell action of the regularized 4D EGB theory recovers the action of the novel 4D EGB theory. Besides, the trace of the field equations has the same form as the one in the novel 4D EGB theory~\cite{Fernandes2,Easson1}. Thus the novel 4D EGB theory may have a hidden scalar degree of freedom. This feature indicates that the novel 4D EGB theory and its regularized counterpart may give the same contributions on the 4D local dynamics. One then is required to check the equivalence of the solutions of the metric and matter fields between these two theories. Mathematically, it is to test whether the solutions in the novel 4D EGB theory satisfy the on-shell condition in the regularized 4D EGB theory. The equivalence of the solutions between two theories has been proved in the static spherically symmetric spacetime~\cite{Lu1,Yang2}. However, such equivalence is broken down in the Taub-NUT metric and the rotating black hole metrics constructed by the Newmann-Janis algorithm~\cite{Hennigar1}. So it indicates that solutions obtained in the novel 4D EGB theory may be different from the ones in the regularized 4D EGB theory in a more complicated spacetime than spherically symmetric one. Thus, it is important to check whether it holds true for other kinds of the spacetime beyond spherical symmetry.

In this paper, we will test the on-shell condition in a particular kind of static cylindrically symmetric spacetime, which is less symmetric than the static spherically symmetric spacetime. We will show that the on-shell condition holds in the regularized 4D EGB gravity and the cylindrically symmetric solution solves both the two theories simultaneously. Further, as a specific example of the cylindrically symmetric metric, we consider a cosmic string. Then, we will solve the higher-dimensional EGB theory, and obtain three sets of cosmic string solutions in the novel 4D EGB theory in the limit $D\rightarrow4$ after the redefinition $\alpha\rightarrow\alpha/(D-4)$~\footnote{It was pointed out in Ref.~\cite{Ai1} that such a continuous procedure contradicts to the discrete behavior of the indices. In this paper we will ignore this index problem and take benefit of the dimensional regularization.}. The on-shell conditions of its regularized counterpart will also be checked again with the three cosmic string solutions. Our result shows that the novel 4D EGB also makes sense on the cylindrically symmetric spacetime.

As a byproduct of this paper, we also investigate the nontrivial effects of the GB term on the cosmic strings. It is well known that topological defects could arise during phase transitions when spontaneous symmetry breaking happens~\cite{Kirzhnits1}.
In the early Universe, these defects are the so-called cosmic strings~\cite{Vilenkin1}. The string tension is an important property, since it is related to the energy scale of the associated phase transition. The tension is usually parameterized by $G_{4}\,\rho$, where $G_{4}$ is the four-dimensional Newton's gravitational constant and $\rho$ the energy density of the string. Our result reveals the nontrivial effects from the GB term on the string mass density. It may then help us to understand how the novel 4D EGB theory and its regularized counterpart modify the energy scale of the associated phase transition.

This paper is arranged as follows. In Sec.~\ref{sec2}, the equivalence of the two theories has been proved in a particular kind of static cylindrically symmetric spacetime. In Sec.~\ref{sec3}, as some specific examples, three sets of cosmic string solutions are given and string mass density is discussed.  A brief conclusion is given in Sec.~\ref{sec4}.

\section{Equivalence of the original and regularized EGB theories}\label{sec2}

\subsection{Field equations of the novel 4D EGB theory}

The action of the $D$-dimensional EGB gravity is
\begin{equation}
	S=\frac{1}{2\kappa^{2}}\int d^{D}x\sqrt{-g}\big(R+\alpha\mathcal{G}\big)+S_{m},
\end{equation}
where $\kappa^{2}=8\pi G_{D}$ with $G_{D}$ the $D$-dimensional Newton's gravitational constant, which will be set to $2\kappa^{2}=1$ for later convenience, $\alpha$ is the GB coupling constant, and $S_{m}$ is the matter action. Note that the cosmological term has been included in the matter part. The GB term $\mathcal{G}$ is defined as
\begin{equation}\label{GB1}
	\mathcal{G}=R^{2}-4R_{KL}R^{KL}+R^{KLPQ}R_{KLPQ}.
\end{equation}
The field equations are given by
\begin{eqnarray}\label{feq1}
	R_{MN}-\frac{1}{2}g_{MN}R+\mathcal{G}_{MN}=\frac{1}{2} T_{MN},
\end{eqnarray}
where $T_{MN}\equiv-\frac{2}{\sqrt{-g}}\frac{\delta S_{m}}{\delta g^{MN}}$ is the energy-momentum tensor and $\mathcal{G}_{MN}$ a tensor carrying the contributions from the GB term:
\begin{eqnarray}\label{gbt1}
	\mathcal{G}_{MN}\!&\!\equiv\!&\!\alpha\Big[2RR_{MN}-4R^{K}_{~M}R_{KN}+2g_{MK}R^{KLPQ}R_{PQNL}\nonumber\\
	&~&-4\,g_{MK}g_{NL}R^{KPLQ}R_{QP}-\frac{1}{2}g_{MN}\mathcal{G}\Big].
\end{eqnarray}
For the case of $D=4$, the GB term~\eqref{GB1} is a topological invariant, and thus it does not affect the Einstein's equations. While it gives nontrivial contributions when the dimension of spacetime is larger than four. Remarkably, many nontrivial effects from the GB term in four dimensions were revealed by redefining the GB coupling constant as $\alpha\rightarrow\alpha/(D-4)$ and taking the limit $D\rightarrow4$ (for detail see e.g. Refs.~\cite{Tomozawa2011,Cognola2013,Glavan1}). In what follows, we will show how this method works for the case of a cosmic string and then investigate contributions of the GB term on the spacetime geometry and characteristics of the cosmic string.

After the theory proposed by Kaluza and Klein~\cite{Kaluza1,Klein1,Klein2}, some recent higher-dimensional theories assume that our Universe is a four-dimensional manifold $\mathcal{M}_{4}$ embedded in a higher-dimensional spacetime (see e.g. Refs.~\cite{Dvali1,Deffayet1,Randall1,Randall2}). Ordinary particles and interactions introduced in the Standard Model of particle physics are confined on $\mathcal{M}_{4}$, while gravity could propagate through the bulk. For a five-dimensional spacetime, $\mathcal{M}_{4}$ is called as a brane, while it is named as a string in six dimensions. By analogy with the latter case, one could use a cylindrically symmetric metric to construct a string-like slice $\mathcal{M}_{d}$ in a $D$-dimensional spacetime with $D=d+2\ge3$:
\begin{equation}
	 ds^{2}_{D}=F(r,\theta)h_{\mu\nu}dx^{\mu}dx^{\nu}+g_{rr}dr^{2}+g_{\theta\theta}dx^{\theta}x^{\theta},
\end{equation}
where $h_{\mu\nu}$ is the induced metric on the manifold $\mathcal{M}_{d}$. For the cosmic string solutions in higher-dimensional spacetime, one could see  Refs.~\cite{Olasagasti1,Gherghetta1,Oda1,Dzhunushaliev1} as examples.
For the usual 4D cosmic string in the early Universe, one could obtain the corresponding metric by fixing the dimension of the manifold $\mathcal{M}_{d}$ to $d=2$.

In this paper, we will start from a $D$-dimensional static cylindrically symmetric spacetime with the metric ansatz given by
\begin{equation}\label{metric1}
	ds^{2}_{D}=H(r)\eta_{\mu\nu}dx^{\mu}dx^{\nu}+dr^{2}+W(r)^{2}d\theta^{2},
\end{equation}
where $\eta_{\mu\nu}$ is the induced metric of a $(D-2)$-dimensional Minkowski manifold with $\mu,\nu=0,1,2,3,5,\cdots,D-2$. When $D=6$, $r$ and $\theta$ become the extra spatial dimensions and our Universe can be viewed as a four-dimensional Minkowski string embedded in a six-dimensional spacetime with $H(r)$ the warped factor of the string. When $D>6$, the dimensions denoted by $x^{i}$ ($i = 5,6,\cdots,D-2$) should be compactified in this braneworld model~\footnote{Note that the critical spacetime dimension of the EGB theory is $D=5$, $6$~\cite{Dadhich1}. For a Lovelock theory with $D>6$, the higher order Lovelock polynomial should be included. Here we only keep the GB term by regarding the theory with $D>6$ as a higher dimensional generalization of the EGB theory.}. When $D=4$, the metric~\eqref{metric1} could be expressed in the cylindrical coordinates $(t,z,r,\theta)$:
\begin{equation}\label{cc1}
	ds^{2}_{4}=-H(r)dt^{2}+H(r)dz^{2}+dr^{2}+W(r)^{2}d\theta^{2},
\end{equation}
where $z$ and $r$ are the axial coordinate parallel to the string and the radial distance, respectively. The string is located at $r=0$. The angular coordinate $\theta$ is in the range $0\leq\theta\leq2\pi$.

By taking into account the metric~\eqref{metric1}, one could obtain the connection and  Riemann tensor as follows:

\begin{strip}
\begin{eqnarray}
	 \Gamma^{K}_{~MN}&=&\delta^{K}_{~~\theta}\,\delta^{r}_{(M}\,\delta^{\theta}_{~N)}\frac{2W'}{W}+\delta^{K}_{~~\beta}\,\delta^{r}_{(M}\,\delta^{\beta}_{~N)}\frac{H'}{H}-\delta^{\theta}_{~M}\,\delta^{K}_{~~r}\,\delta^{\theta}_{~N}WW'-\delta^{\alpha}_{~M}\,\delta^{K}_{~~r}\,\delta^{\beta}_{~N}\eta_{\alpha\beta}\frac{H'}{2},\\
	 R^{K}_{~LMN}&=&\delta^{K}_{~~\theta}\,\delta^{r}_{~L}\,\delta^{r}_{[M}\delta^{\theta}_{~N]}\frac{2W''}{W}+2\delta^{K}_{~~r}\,\delta^{\theta}_{~L}\,\delta^{\theta}_{[M}\,\delta^{r}_{~N]}WW''+2\,\delta^{K}_{~~\alpha}\,\delta^{r}_{~L}\,\delta^{r}_{[M}\,\delta^{\alpha}_{~N]}\bigg[\frac{H''}{2H}-\Big(\frac{H'}{2H}\Big)^{2}\bigg]\nonumber\\
	 &~&+2\,\delta^{K}_{~~r}\,\delta^{\beta}_{~L}\,\delta^{\alpha}_{[M}\,\delta^{r}_{~N]}\eta_{\alpha\beta}\bigg[\frac{H''}{2H}-\Big(\frac{H'}{2H}\Big)^{2}\bigg]H+\delta^{K}_{~~\alpha}\,\delta^{\theta}_{~L}\,\delta^{\theta}_{[M}\,\delta^{\alpha}_{~N]}\frac{WW'H'}{H}\nonumber\\
	 &~&-\delta^{K}_{~~\theta}\,\delta^{\beta}_{~L}\,\delta^{\theta}_{[M}\,\delta^{\alpha}_{~N]}\eta_{\alpha\beta}\frac{W'H'}{W}+\delta^{K}_{~~\rho}\,\delta^{\beta}_{~L}\,\delta^{\alpha}_{[M}\,\delta^{\rho}_{~N]}\eta_{\alpha\beta}\frac{(H')^{2}}{2H},\label{RKLMN1}\\
	 R_{PQNL}&=&2\delta^{\theta}_{~P}\delta^{r}_{~Q}\delta^{r}_{[N}\delta^{\theta}_{~L]}WW''+2\delta^{r}_{~P}\delta^{\theta}_{~Q}\delta^{\theta}_{[N}\delta^{r}_{~L]}WW''+\delta^{\beta}_{~P}\,\delta^{r}_{~Q}\,\delta^{r}_{[N}\delta^{\alpha}_{~L]}\eta_{\alpha\beta}\frac{H}{2}\bigg[\frac{2H''}{H}-\Big(\frac{H'}{H}\Big)^{2}\bigg]\nonumber\\
	 &~&+\delta^{r}_{~P}\,\delta^{\beta}_{~Q}\,\delta^{\alpha}_{[N}\delta^{r}_{~L]}\eta_{\alpha\beta}\frac{H}{2}\bigg[\frac{2H''}{H}-\Big(\frac{H'}{H}\Big)^{2}\bigg]+\delta^{\beta}_{~P}\,\delta^{\theta}_{~Q}\,\delta^{\theta}_{[N}\delta^{\alpha}_{~L]}\eta_{\alpha\beta}WW'H'\nonumber\\
	 &~&+\delta^{\sigma}_{~P}\,\delta^{\beta}_{~Q}\,\delta^{\alpha}_{[N}\delta^{\rho}_{~L]}\eta_{\rho\sigma}\eta_{\alpha\beta}\frac{(H')^{2}}{2}-\delta^{\theta}_{~P}\,\delta^{\beta}_{Q}\,\delta^{\theta}_{[N}\delta^{\alpha}_{~L]}\eta_{\alpha\beta}WW'H',\\
	 R_{NL}&=&\delta^{r}_{~N}\,\delta^{r}_{~L}\Big[\frac{D-2}{4}\Big(\frac{H'}{H}\Big)^{2}-\frac{D-2}{2}\frac{H''}{H}-\frac{W''}{W}\Big]-\delta^{\alpha}_{~N}\,\delta^{\beta}_{~L}\eta_{\alpha\beta}\frac{1}{2}\Big[\frac{D-4}{2}\frac{(H')^{2}}{H}+\frac{W'H'}{W}+H''\Big]\nonumber\\
	 &~&-\delta^{\theta}_{~N}\,\delta^{\theta}_{~L}W\Big[\frac{D-2}{2}\frac{W'H'}{H}+W''\Big],\label{RMN1}
\end{eqnarray}
where the prime denotes the derivative with respect to the coordinate $r$. The round or square bracket in subscript represents the symmetrization or antisymmetrization of the indices. A straightforward calculation gives the nonvanishing components of the gravitational field equations:
\begin{subequations}\label{feq2}
 \begin{align}
  &(\mu,\nu):& \frac{T_{tt}}{H}=&\frac{\alpha_{2,3,4}}{D-2}\frac{H'}{H}\bigg[
     \frac{ (D-10)(D-5) }{8} \Big(\frac{H'}{H}\Big)^{3}
     +(D-7)\Big(\frac{H'}{H}\Big)^{2}\frac{W'}{W}+(D-5)\frac{H'}{H}\frac{H''}{H}
     +2\frac{H'}{H}\frac{W''}{W}\nonumber\\
     &~&&+4\frac{H''}{H}\frac{W'}{W}\bigg]
     -(D-3)\bigg[\frac{D-6}{4}\Big(\frac{H'}{H}\Big)^{2}
     +\frac{H'}{H}\frac{W'}{W}+\frac{H''}{H}
      +\frac{2}{D-3}\frac{W''}{W}\bigg],\label{feq2c1}\\
     &(r,r):& T_{rr}=&
     \alpha_{2,3,4} \Big(\frac{H'}{H}\Big)^{3}
       \bigg[-\frac{(D-5) }{8} \frac{H'}{H}
             -\frac{W'}{W}\bigg]
       +(D-2)\frac{H'}{H}\bigg[\frac{D-3}{4}\frac{H'}{H}
      +\frac{W'}{W}\bigg],\label{feq2c2}\\
        &(\theta,\theta):&\frac{T_{\theta\theta}}{W^2}=&
  - \alpha_{2,3,4}\Big(\frac{H'}{H}\Big)^{2} \bigg[\frac{(D-9)}{8} \Big(\frac{H'}{H}\Big)^{2}
     +\frac{H''}{H}\bigg]+(D-2)\bigg[\frac{D-5}{4}\Big(\frac{H'}{H}\Big)^{2}
      +\frac{H''}{H}\bigg],\label{feq2c3}
 \end{align}
\end{subequations}
\end{strip}
where
\begin{equation}
 \alpha_{2,3,4}=(D-2)(D-3)(D-4)\alpha/2.
\end{equation}
The contributions of the GB term labeled with coupling constant $\alpha$ in each component of the gravitational field equations appear with a factor $(D-4)$. This is a general feature of the $D$-dimensional EGB theory (one could check it with Eq.~(8) in Ref.~\cite{Glavan1}). Therefore, no contribution of the GB term could be found in the 4D EGB theory. While, by introducing the redefinition $\alpha\rightarrow\alpha/(D-4)$~\cite{Glavan1}, all such factors $(D-4)$ in the gravitational field equations are canceled. In the limit $D\rightarrow4$, the resultant field equations read
\begin{subequations}\label{feq4D}
\begin{eqnarray}
	(\mu,\nu):~\frac{T_{tt}}{H}&=&\frac{\alpha}{2}\frac{H'}{H}\bigg[
     \frac{ 3 }{4} \Big(\frac{H'}{H}\Big)^{3}
    -3\Big(\frac{H'}{H}\Big)^{2}\frac{W'}{W}-\frac{H'}{H}\frac{H''}{H}
     \nonumber\\
     &~&+2\frac{H'}{H}\frac{W''}{W}+4\frac{H''}{H}\frac{W'}{W}\bigg]
     +\frac{1}{2}\Big(\frac{H'}{H}\Big)^{2}\nonumber\\
     &~&
     -\frac{H'}{H}\frac{W'}{W}-\frac{H''}{H}
      -2\frac{W''}{W},\label{feq2c1}\\
   	(r,r):~ T_{rr}&=&
    \alpha\Big(\frac{H'}{H}\Big)^{3}
       \bigg[\frac{H'}{8H}
             -\frac{W'}{W}\bigg]
       +\frac{1}{2}\Big(\frac{H'}{H}\Big)^{2}
     \nonumber\\
     &~& +2\frac{H'}{H}\frac{W'}{W},\label{feq2c2}\\
       (\theta,\theta):~\frac{T_{\theta\theta}}{W^2}&=&
  \alpha\Big(\frac{H'}{H}\Big)^{2} \bigg[\frac{5}{8} \Big(\frac{H'}{H}\Big)^{2}
     -\frac{H''}{H}\bigg]-\frac{1}{2}\Big(\frac{H'}{H}\Big)^{2}
     \nonumber\\
     &~& +2\frac{H''}{H}.\label{feq2c3}
 \end{eqnarray}
\end{subequations}

Thus the theory has well-defined field equations in $D\to4$ limit for the static cylindrically symmetric spacetime. Obviously, nontrivial contribution from the GB term appears in the four-dimensional field equations~\eqref{feq4D}.

\subsection{Field equations of the regularized 4D EGB theory}

Although the well-defined field equations in $D\to4$ limit for the particular kind of static cylindrically symmetric spacetime~\eqref{metric1} can be obtained after the redefinition $\alpha\rightarrow\alpha/(D-4)$, generally, the 4D EGB gravity would admit neither an intrinsically four-dimensional definition nor well-defined four-dimensional field equations~\cite{Gurses1,Fernandes2}. To address this problem, a regularized 4D EGB theory was constructed~\cite{Fernandes2,Hennigar1}. The main mechanism of such theory is to introduce an extra GB term constructed from the conformal metric,
\begin{equation}
 \tilde{g}_{MN}=e^{2\psi(r)}g_{MN},	
\end{equation}
to cancel the original GB term. The resultant gravitational part of the action contains an Einstein-Hilbert term and some extra terms involving $\psi(r)$:
\begin{eqnarray}\label{r4DEGBa1}
	S_{g}&=&\int d^{4}x\sqrt{-g}\Big[R+\alpha\big(4G^{MN}\nabla_{M}\psi\nabla_{N}\psi-\psi\,\mathcal{G}\nonumber\\
	&~&+4\Box\psi\nabla_{K}\psi\nabla^{K}\psi+2\nabla_{K}\psi\nabla_{L}\psi\nabla^{K}\psi\nabla^{L}\psi\big)\Big],
\end{eqnarray}
The field equations of this theory were already obtained in Refs.~\cite{Fernandes2,Hennigar1}. It is worth to note that the extra GB term vanishes when the scalar field $\psi$ is on-shell. In other words, the on-shell action of the regularized 4D EGB theory equals to the one of the original novel 4D EGB theory. And when the system is on-shell, one could investigate the novel 4D EGB theory through the field equations of the regularized one. A natural question then arises as to check whether the spacetime geometry given in the novel 4D EGB theory satisfies the on-shell condition of the regularized 4D EGB theory. For a spherically symmetric spacetime, the equivalence of the actions in the regularized 4D EGB theory and in the original one has been proved (see e.g. Refs.~\cite{Lu1,Yang2}). In this section, we will use our cosmic string solutions to test whether the actions of these two theories are still equivalent when the symmetry of spacetime is beyond the spherical symmetry. Our result supports that the  regularized 4D EGB theory is equivalent to the original one in a specific cylindrically symmetric spacetime. Particularly, we also prove the equivalence in a more general case. To give a short conclusion here, we shall say that when the metric ansatz follows our assumption~\eqref{cc1}, the novel 4D EGB theory and its regularized counterpart are equivalent.

The gravitational field equations of the regularized 4D EGB theory are
\begin{equation}
	G_{MN}=\alpha \mathcal{H}_{MN}+\frac{1}{2}T_{MN},
\end{equation}
where $G_{MN}$ is the Einstein tensor and $\mathcal{H}_{MN}$ is given by~\cite{Fernandes2}

\begin{strip}
\begin{eqnarray}
\mathcal{H}_{MN}\!&\!=\!&\!2R\left(\nabla_\mu\nabla_\nu \psi-\nabla_\mu\psi\nabla_\nu\psi\right)+2G_{\mu\nu}\left[(\nabla\psi)^2-2\Box\psi\right]+4G_{\nu\alpha}\left(\nabla^\alpha\nabla_\mu\psi-\nabla^\alpha\psi\nabla_\mu\psi\right)\nonumber\\
&&+4G_{\mu\alpha}\left(\nabla^\alpha\nabla_\nu\psi-\nabla^\alpha\psi\nabla_\nu\psi\right)
+4R_{\mu\alpha\nu\beta}\left(\nabla^\beta\nabla^\alpha\psi-\nabla^\alpha\psi\nabla^\beta\psi\right)+4\nabla_\alpha\nabla_\nu\psi\left(\nabla^\alpha\psi\nabla_\mu\psi-\nabla^\alpha\nabla_\mu\psi\right)\nonumber\\
&&+4\nabla_\alpha\nabla_\mu\psi\nabla^\alpha\psi\nabla_\nu\psi+4\Box\psi\nabla_\nu\nabla_\mu\psi
-4\nabla_\mu\psi\nabla_\nu\psi\left[(\nabla\psi)^2+\Box\psi\right]^2-g_{\mu\nu}\Big\{2R\left[\Box\psi-(\nabla\psi)^2\right]\nonumber\\
&&+4G^{\alpha\beta}\left(\nabla_\beta\nabla_\alpha\psi-\nabla_\alpha\psi\nabla_\beta\psi\right)+2(\Box\psi)^2-(\nabla\psi)^4+2\nabla_\beta\nabla_\alpha\psi\left(2\nabla^\alpha\psi\nabla^\beta\psi-\nabla^\beta\nabla^\alpha\psi\right)\Big\}.
\end{eqnarray}
The field equation of the scalar field $\psi$ reads:
\begin{eqnarray}\label{r4EGBgfeq1}
	\frac{1}{8}\mathcal{G}&=&R^{KL}\nabla_{K}\psi\nabla_{L}\psi-G^{KL}\nabla_{K}\nabla_{L}\psi-\Box\psi\nabla_{K}\psi\nabla^{K}\psi+\nabla_{K}\nabla_{L}\psi\nabla^{K}\nabla^{L}\psi-(\Box\psi)^{2}-2\nabla_{K}\psi\nabla_{L}\psi\nabla^{K}\nabla^{L}\psi.
\end{eqnarray}
With the cylindrical coordinates~\eqref{cc1}, a straightforward calculation gives the following nonvanishing components of the gravitational field equations:
\begin{subequations}\label{r4DEGBgfqf}
	\begin{eqnarray}
		\frac{T_{tt}}{H}&=&\frac{1}{2}\Big(\frac{H'}{H}\Big)^{2}-\frac{H''}{H}-\frac{2W''}{W}-\frac{H'}{H}\frac{W'}{W}-2\alpha\bigg[\frac{2 H''}{H}\frac{W'}{W}+\frac{2 H' }{H}\frac{W''}{W}-\Big(\frac{H'}{H}\Big)^2\frac{ W'}{W}\bigg]\varphi\nonumber\\
		&~&-\alpha\bigg[\frac{2 H''}{H}+\frac{2 H'}{H }\frac{W'}{W}-\Big(\frac{H'}{H}\Big)^2+\frac{4 W''}{W}\bigg]\varphi^{2}+2 \alpha  \varphi ^4-4\alpha\bigg[2 \varphi ^2+  \Big(\frac{H'}{H}+\frac{2 W'}{W}\Big)\varphi+\frac{H'}{H }\frac{W'}{W}\bigg]\varphi',\label{r4DEGBgfq1}\\
		T_{rr}&=&\frac{2 H' }{ W}\frac{4 W'}{ W}+\frac{1}{2}\Big(\frac{H'}{H}\Big)^2+6 \alpha  \Big(\frac{H'}{H}\Big)^2 \frac{  W'}{W}\varphi+3 \alpha  \Big(\frac{H'}{H}+\frac{4 W'}{W}\Big)\frac{  H'}{H} \varphi^{2}+8\alpha\Big(\frac{H'}{H}+\frac{W'}{W}\Big)\varphi^{3}+6\alpha\varphi^{4},\label{r4DEGBgfq2}\\
		\frac{T_{\theta\theta}}{W^2}&=&\frac{2 H''}{H}-\frac{1}{2}\Big(\frac{H'}{H}\Big)^2-2\alpha\bigg[\Big(\frac{H'}{H}\Big)^2-\frac{2 H''}{H}\bigg]\frac{H'}{H}\varphi-\alpha\bigg[\Big(\frac{H'}{H}\Big)^2-\frac{4 H''}{H}\bigg]\varphi^{2}-2\alpha\varphi^{4}+2\alpha\bigg[4\varphi^{2}+4\frac{H'}{H}\varphi+\Big(\frac{H'}{H}\Big)^2\bigg]\varphi',\nonumber\\
		&~&\label{r4DEGBgfq3}
	\end{eqnarray}
\end{subequations}
\end{strip}
where $\varphi$ is defined as $\varphi\equiv\psi'(r)$. For a given cylindrical solution in the novel 4D EGB theory, the above equations correspond to three constraints on the scalar field $\varphi$. The equivalence of the novel 4D EGB theory and its regularized counterpart then requires that the resultant constraints should meet with the on-shell condition~\eqref{r4EGBgfeq1}.

\subsection{Equivalence of the two theories}

As we have mentioned before, if the on-shell action of the regularized 4D EGB theory equals to the one of the novel 4D EGB theory in a static cylindrically symmetric spacetime~\eqref{cc1}, both of them will give the same solutions. From Eq.~\eqref{r4EGBgfeq1}, we find that the on-shell condition does not involve the matter content. Thus it is convenient to eliminate the matter part in the gravitational field equations~\eqref{r4DEGBgfqf} with Eq.~\eqref{feq4D}. The resultant field equations read
\begin{subequations}\label{r4EGBgfeq3}
	\begin{eqnarray}
		0\!&\!=\!&\!\bigg[ \Big(\frac{3}{4}\frac{H'}{H}+\frac{W'}{W} \Big)\frac{H'}{H}-\frac{H''}{H}+2\frac{W''}{W}+\Big(\frac{  H'}{H}-\varphi\Big)\varphi\nonumber\\
		\!&\!~\!&\!-2 \varphi '\bigg]X+2\frac{W'}{W}X'+(2\varphi X )',\\
		0\!&\!=\!&\!\Big(\frac{2 W'}{W}-\frac{H'}{4 H}+\frac{3 \varphi }{2}\Big)X^{3/2},\\
		0\!&\!=\!&\!\bigg[\frac{5}{4} \Big(\frac{H'}{H}\Big)^2-\frac{2 H''}{H}-\Big(\frac{  H'}{H}-\varphi\Big)\varphi-4 \varphi '\bigg]X,
	\end{eqnarray}
\end{subequations}
with
\begin{equation}\label{X1}
	X=\Big(\frac{H'}{H}+2 \varphi\Big)^{2}.
\end{equation}
One could check that the above field equations~\eqref{r4EGBgfeq3} are identities when $X$ equals to zero. From the definition~\eqref{X1}, we know that it requires the scalar field to satisfy
\begin{equation}\label{onshell1}
	\varphi=-\frac{H'}{2H}.
\end{equation}
On the other hand, the on-shell condition~\eqref{r4EGBgfeq1} could be expressed as follows:
\begin{eqnarray}\label{r4DEGBsfq2}
	0&=&\Big(2 \varphi+\frac{H'}{H}\Big)\bigg\{\frac{2 H'' }{H }\frac{ W'}{W}+\frac{H'}{H}\frac{W''}{W}-\Big(\frac{H'}{H}\Big)^3\nonumber\\
	&~&+\bigg[\frac{H''}{H}+\frac{3 H' }{H }+\frac{W'}{W}-\Big(\frac{H'}{H}\Big)^2+\frac{2 W''}{W}+6\bigg]\varphi\nonumber\\
	&~&2\Big(\frac{H'}{H}+\frac{W'}{W}\Big)\varphi^{2}+\Big(\frac{H'}{H}+\frac{4 W'}{W}\Big)\varphi'^{2}\bigg\}.
\end{eqnarray}
Obviously, the above scalar field equation is an identity when the scalar field $\varphi$ satisfies the relation~\eqref{onshell1}. Indeed, a straightforward proof could be given by substituting this expression into the field equations~\eqref{r4DEGBgfqf} to eliminate $\varphi$. One could check that the resultant field equations just recover the ones~\eqref{feq4D} of the novel 4D EGB theory. Thus, we can conclude that the two 4D EGB theories are equivalent in the static cylindrically symmetric spacetime~\eqref{cc1}. In other words, solutions of the static cylindrically symmetric metric~\eqref{cc1} of the novel 4D EGB theory are also solutions of the regularized one.

\section{Cosmic String Solutions}\label{sec3}

As a specific example of the static cylindrically symmetric metic~(\ref{metric1}), we consider the cylindrical cosmic string.
Cosmic strings have various astrophysical phenomena~\cite{Hogan1,Vachaspati1,Bennett1,Caldwell1,Kaiser1,Vachaspati2,Danos1,Brandenberger1,Vilenkin2,Jusufi1,Ovgun1,Yamauchi1}. For example, the formation of a network of cosmic strings is expected to generate gravitational waves with a large amplitude~\cite{Hogan1,Vachaspati1,Bennett1,Caldwell1}. Cosmic strings could affect the cosmic microwave background photons and therefore give rise to distinctive signatures on the cosmic microwave background~\cite{Kaiser1,Vachaspati2,Danos1,Brandenberger1}. They could also bend the light curves that result in gravitational lensings (see e.g. Refs.~\cite{Vilenkin2,Jusufi1,Ovgun1,Yamauchi1}). Other researches can be found in Refs.~\cite{Hammond1,Blanco-Pillado1,Ringeval1,Sadr1}.

An important property of a cosmic string is its tension. As is well known, the cosmic string is a kind of topological defect in the early Universe~\cite{Vilenkin1}, while topological defects usually arise during phase transitions when spontaneous symmetry breaking happens. Their formations require a sufficient high temperature $T>T_{c}=\sqrt{6}\eta$ with $T_{c}$ the critical temperature in the second phase transition and $\eta$ the energy breaking scale~\cite{Kirzhnits1}. Thus the string tension is related to the energy scale of the associated phase transitions. According to the astrophysical phenomena of cosmic strings, the string tension could be constrained by the observations. The cosmic microwave background data from the Planck Satellite constrains the tension of the Nambo-Goto string to $G_{4}\,\rho\lesssim10^{-7}$~\cite{Ade1}. While the LIGO and Virgo collaboration has limited the tension of the Nambu-Goto string as $G_{4}\,\rho<2\times10^{-14}$~\cite{Abbott1,Abbott2}.

In the novel 4D EGB theory, the above constrains will change when the GB term  contributes a nontrivial effect to the string mass density. In this section, we will give three sets of solutions and show such nontrivial effects from the GB term. We will then test the on-shell condition of the regularized 4D EGB theory with these solutions. The result shows that they are also the string solutions of the regularized 4D EGB theory, and hence supports the conclusion obtained in the last section. Thus the nontrivial effects found in this section may help us to understand how the novel 4D EGB theory and its regularized counterpart affect the energy scale of the phase transition.

For a cosmic string, we consider the following matter action of a scalar field:
\begin{equation}
	S_{m}=\int d^{D}x\sqrt{-g}\Big[-\frac{1}{2}g^{KL}\nabla_{K}\phi(r)\nabla_{L}\phi(r)-V(\phi)\Big],
\end{equation}
for which the energy-momentum tensor is given by
\begin{eqnarray}\label{emt1}
	T_{MN} 
	 =\nabla_{M}\phi\nabla_{N}\phi-\frac{1}{2}g_{MN}\nabla^{L}\phi\nabla_{L}\phi 
	-g_{MN}V(\phi).
\end{eqnarray}
The metric ansatz for the cosmic string in $D$-dimensional spacetime is given by (\ref{metric1}), with which the independent equations can be simplified to

\begin{strip}
\begin{subequations}\label{feq3}
\begin{eqnarray}
	0
	&=& {\alpha_{2,3,4}}\frac{H'}{H} \Bigg[
	 (D-8)\Big(\frac{H'}{H}\Big)^{3}
     -{2(D-7)}\Big(\frac{H'}{H}\Big)^{2}\frac{W'}{W}
	+6\frac{H'}{H}\frac{H''}{H}
    -4\frac{H'}{H}\frac{W''}{W}
    -8\frac{H''}{H}\frac{W'}{W}
        \Bigg]
		\nonumber\\
	&~& -(D-2) \Bigg[
         2\frac{H''}{H}
         +(D-4)\Big(\frac{H'}{H}\Big)^{2}
	     -4\frac{W''}{W}
	     -2(D-3)\frac{H'}{H}\frac{W'}{W} \Bigg] ,\label{feq3c1}\\
 \phi'^{2}
	&=& \alpha_{2,3,4}\Big(\frac{H'}{H}\Big)^{2} \Bigg[
	     \frac{H''}{H}
	     -\frac{1}{2}\Big(\frac{H'}{H}\Big)^{2}
	     -\frac{H'}{H}\frac{W'}{W} \Bigg]
      -(D-2)  \Bigg[\frac{H''}{H}-
         \frac{1}{2}\Big(\frac{H'}{H}\Big)^{2}
	        -\frac{H'}{H}\frac{W'}{W} \Bigg] ,\label{feq3c2}\\
2V
	&=& \alpha_{2,3,4} \Big(\frac{H'}{H}\Big)^{2}\Bigg[
	        \frac{H''}{H}+
	         \frac{D-7}{4}\Big(\frac{H'}{H}\Big)^{2}
            +\frac{H'}{H}\frac{W'}{W}  \Bigg]
	-(D-2) \Bigg[\frac{H''}{H}
            +\frac{D-4}{2}\Big(\frac{H'}{H}\Big)^{2}
            +\frac{H'}{H}\frac{W'}{W}\Bigg].  \label{feq3c3}
\end{eqnarray}
\end{subequations}
\end{strip}
We now have three independent field equations while with four functions, $H$, $W$, $\phi$, and $V$. Therefore, if one of the functions is fixed, we can determine the solution of the other functions. On the other hand, we find that Eq.~\eqref{feq3c1} gives a constraint between the functions $H$ and $W$, with which $H$ could be expressed as a function of $W$ and vise versa. By analyzing Eqs.~\eqref{feq3c2} and~\eqref{feq3c3}, we could further solve the rest quantities $\phi$ and $V$ as functions of either $H$ or $W$. Indeed, Eq.~\eqref{feq3c1} is a second-order linear differential equation for $W$:
\begin{equation}
   W''=\frac{A_{1}}{A_{2}}\frac{H'}{H}W'+\frac{B}{A_{2}}W,\label{W1''}
\end{equation}
where
\begin{subequations}
 \begin{eqnarray}
     A_{1}&=&
     \frac{D-3}{2 }-\alpha_{3,4}\bigg[\frac{D-7}{4} \Big(\frac{H'}{H}\Big)^2+\frac{ H''}{H}\bigg],\nonumber\\
     &~&\\
     A_{2}&=&\frac{\alpha_{3,4}}{2}    \Big(\frac{H'}{H}\Big)^2-1\,,\\
     B&=&
     \frac{\alpha_{3,4}}{4}\Big(\frac{H'}{H}\Big)^{2}\bigg[\frac{D-8}{2}\Big(\frac{H'}{H}\Big)^{2}+3\frac{H''}{H}\bigg]\nonumber\\
     &~&
     -\frac{1}{2} \bigg[\frac{H''}{H}
     +\frac{D-4}{2}\Big(\frac{H'}{H}\Big)^{2}\bigg]
     ,
    \end{eqnarray}
\end{subequations}
with  $\alpha_{3,4}= (D-3)(D-4)\alpha$.
Obviously, Eq.~\eqref{feq3c1} and therefore Eqs.~\eqref{feq3c2} and~\eqref{feq3c3} can be solved by assuming some specific forms of $H$. In the following, we will give three sets of string solutions in a $D$-dimensional spacetime. Besides, we will introduce the redefinition $\alpha\rightarrow\alpha/(D-4)$ in each set of our solutions, such that the nontrivial contributions of the GB term on the characters of the cosmic strings may appear in the limit $D\rightarrow4$. Note that the spacetime dimension in our solutions is not fixed. Thus our solutions are also useful to construct braneworld models.

\subsection{$H(r)=e^{-kr}$}

We first assume that the metric component $H$ obeys
\begin{equation}
	H=e^{-kr},\label{H1}
\end{equation}
where $k$ is a positive parameter. As shown in the previous context, the solution of $W$ could be obtained by taking into account the field equation~\eqref{W1''}:
\begin{equation}
	W=e^{- k r/2}+C_{1} e^{(D-2) k r/2}, \label{W1}
\end{equation}
where $C_{1}$ is an integration constant. Indeed, from Eq.~\eqref{W1''} we will get two integration constants in the solution of $W$. The first one is the factor of the first term on the r.h.s. of Eq.~\eqref{W1}, and the second one is the factor of the second term. With the consideration that the boundary of the spacetime could be maximally symmetric we assume that the first integration constant is nonvanishing, while do not constrain the second one. Obviously, the first integration constant could be absorbed into $\theta$ with the second one changing to $C_{1}$. Again, here and after, the GB coupling constant will be redefined as $\alpha\rightarrow\alpha/(D-4)$ in order to give a nontrivial contribution from the GB term in the limit $D\rightarrow4$.

Having the expressions of $H$ and $W$, one finds that Eqs.~\eqref{feq3c2} and~\eqref{feq3c3} become the linear equations of $\phi'^{2}$ and $V$, respectively. Then the solutions of $\phi'$ and $V$ are obtained as
\begin{eqnarray}
    \phi'\!\!&=&\!\!  \frac{k(D\!-\!1) C_{2}  }{4  }\sqrt{\frac{\Theta(r)-1}{\Theta(r)}},\label{phi1}\\	
      V  \!\!&=&\!\! -\frac{(D\!-\!2) (D\!-\!1) k^2-C_{3}}
                  {4 \Theta(r)}
             \!-\!\frac{C_{3}}{8}
               \frac{\Theta(r)-1}{\Theta(r)},  ~~~~                  \label{V1}
\end{eqnarray}
where
\begin{subequations}
\begin{eqnarray}
    \Theta(r)&=&C_{1}\, e^{(D-1)k r/2}+1\,,\label{Theta1}\\
	C_{2}&=& 2 \sqrt{\frac{D-2}{D-1 } \big[\alpha  (D-3) k^2-2\big]},\label{C31}\\
	C_{3}&=&\alpha  (D-3) (D-2) (D-1) k^4/2.
\end{eqnarray}
\end{subequations}
By integrating Eq.~\eqref{phi1}, we then get the solution of the scalar field:
\begin{equation}
	\phi=C_2 \tanh ^{-1}\big(\sqrt{(\Theta-1)/\Theta}\big)\,.\label{phi2}
\end{equation}
From the definition~\eqref{C31}, we find that the constant $C_{2}$ becomes a complex when the GB coupling constant satisfies $1/\alpha>(D-3) k^2/2$. Thus there is a constraint for this set of solution.  Obviously, a real scalar field solution also requires the integration constant $C_{1}$ to be positive. From the above solution, we could express the scalar potential with respect to the scalar field as follows:
\begin{eqnarray}\label{Vphi1}
	V(\phi)\!&=&\!-\frac{(D\!-\!2) (D\!-\!1) k^2\!-\!C_{3}}
                  {4}\text{sech}^{2}(\phi/C_{2})\nonumber\\
                  &~&
             \!-\frac{C_{3}}{8}\text{tanh}^{2}(\phi/C_{2}).
\end{eqnarray}
The energy density of the scalar field is then given as
\begin{eqnarray}
	\rho(r)&=&-g^{00}T_{00}-\rho_{0}. \label{rho1}  
\end{eqnarray}
Here $\rho_{0}$ is the vacuum energy density of the cosmological constant term included in the scalar potential $V$, and it is given by
\begin{equation}
	\rho_{0}=\frac{C_2^2 (D-1)^2 k^2}{64} -\frac{(D-1) (D-2) k^2}{8},
\end{equation}
which contains the contribution of the GB term labeled by the coupling constant $\alpha$.
Then it is interesting that the energy density of the scalar field is turned out to be zero:
\begin{eqnarray}
	\rho(r)=0.\label{rho1a}
\end{eqnarray}

\begin{figure}[!htb]
\center{
\subfigure[]{\includegraphics[width=3.9cm]{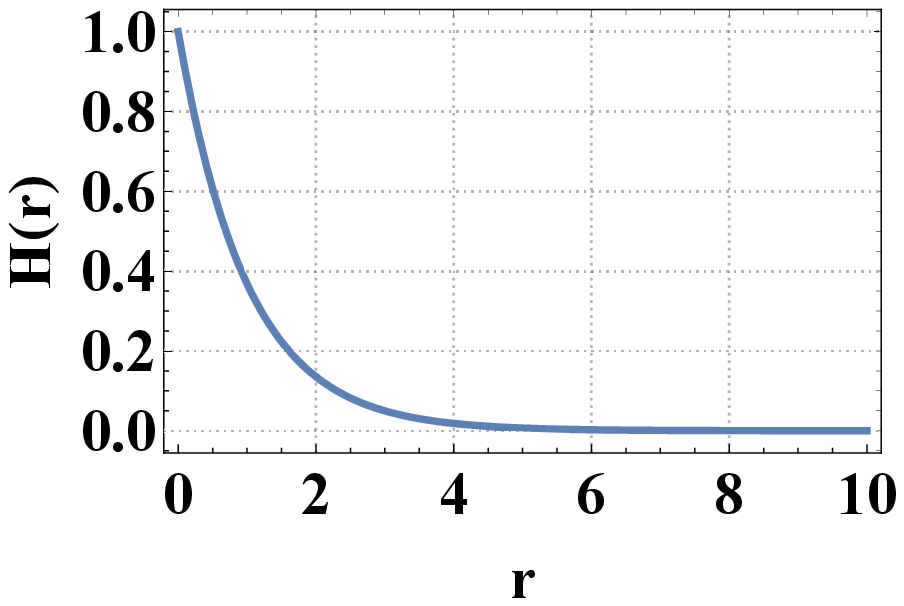}\label{Hc1}}
\quad
\subfigure[]{\includegraphics[width=3.9cm]{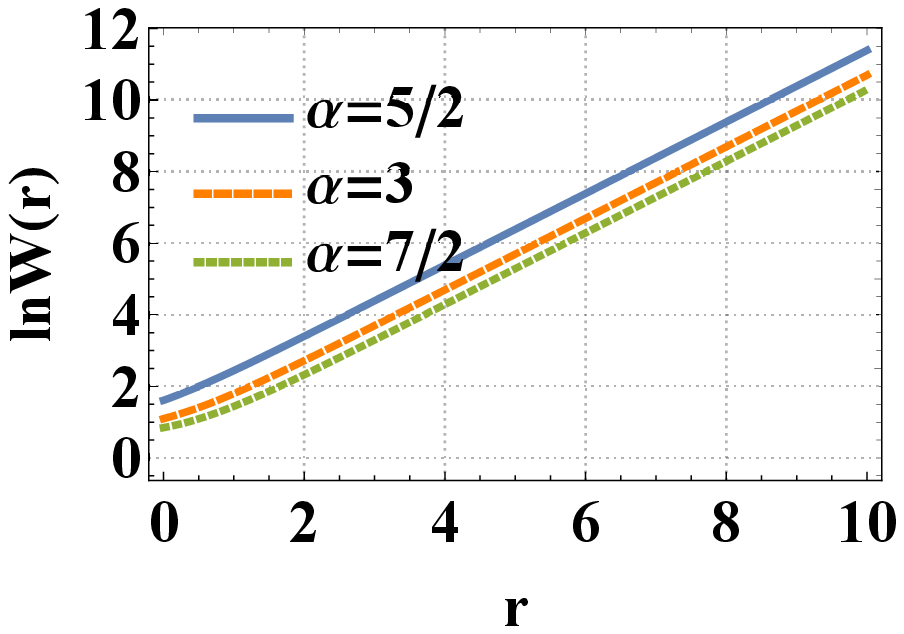}\label{Wc1}}
\quad
\subfigure[]{\includegraphics[width=3.9cm]{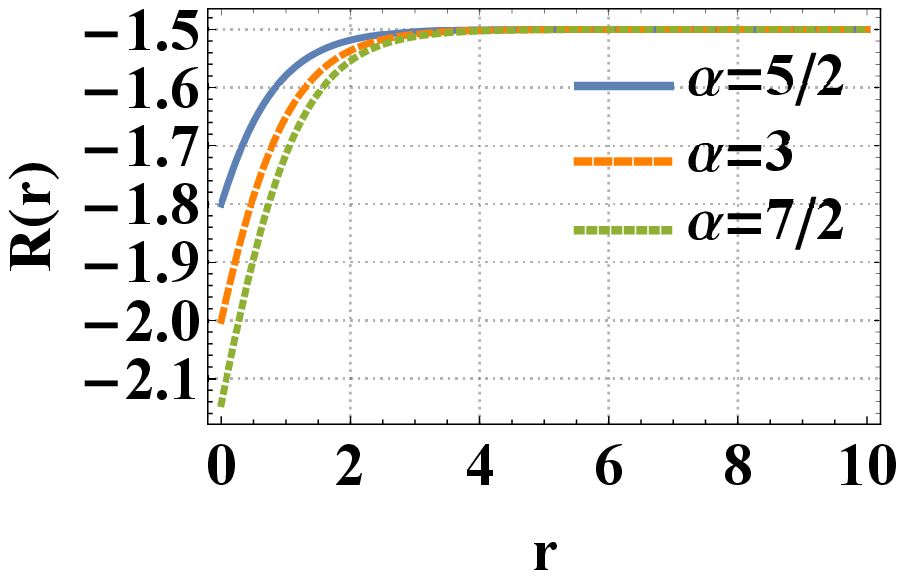}\label{Rc1}}
}
\caption{Plots of the metric components~\eqref{H1}-\eqref{W1}, and the curvature scalar~\eqref{R1} in the limit $D\rightarrow4$ for the first solution. (a) The metric component $H(r)$. (b) The metric component  $W(r)$. (c) The curvature scalar $R(r)$.  The parameters are set to $k=1$ and $C_{1}=-1/(1-\alpha/2)$.}
\label{solution1}
\end{figure}

\begin{figure}[!htb]
\center{
\subfigure[]{\includegraphics[width=3.9cm]{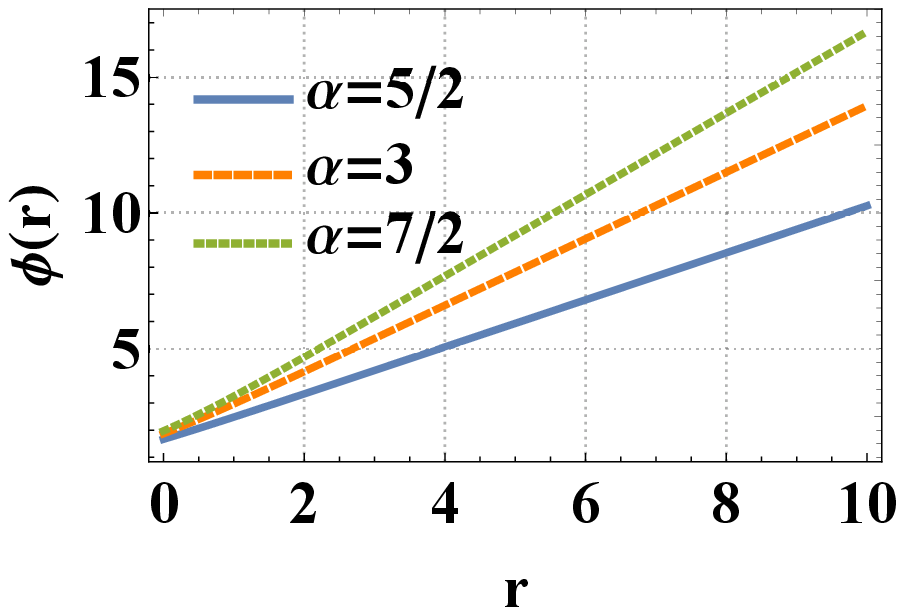}\label{phic1}}
\quad
\subfigure[]{\includegraphics[width=3.9cm]{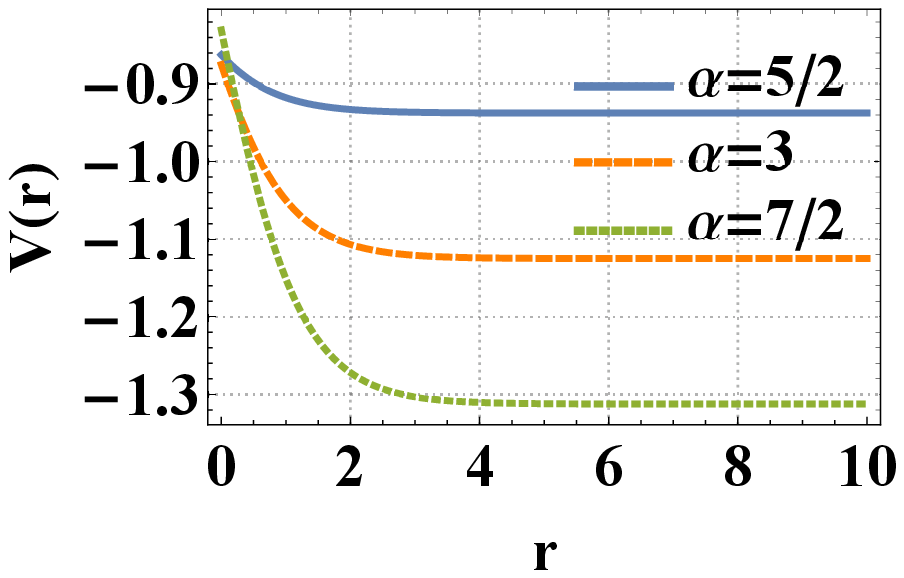}\label{Vc1}}
\quad
\subfigure[]{\includegraphics[width=3.9cm]{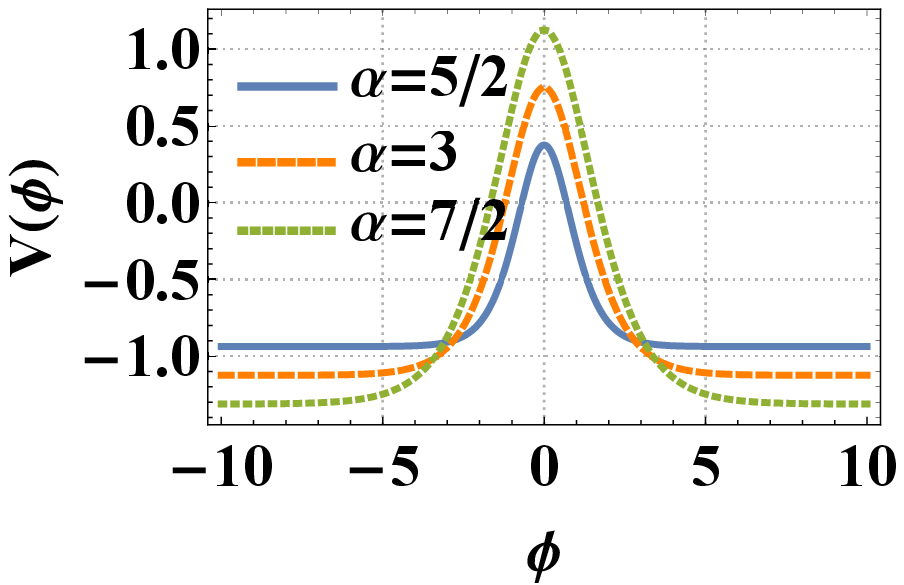}\label{Vphic1}}
}
\caption{The scalar field~\eqref{phi2} and the scalar potential,~\eqref{V1} or~\eqref{Vphi1}, in the limit $D\rightarrow4$ for the first solution. (a) The scalar field $\phi(r)$. (b) The scalar potential $V(r)$. (c) The scalar potential $V(\phi)$ as a function of the scalar field. The parameters are set to $k=1$ and $C_{1}=-1/(1-\alpha/2)$.}
\label{phiV1}
\end{figure}

By taking into account the metric components and the trace of the Ricci curvature~\eqref{RMN1}, we have the following curvature scalar
\begin{equation}
	R(r)=\!-\frac{D-1}{\Theta}\frac{k^2}{2}-\frac{(D-1) (D-2)}{4}  k^2,\label{R1}
\end{equation}
which approaches to $-(D-1) (D-2) k^2/4$ for $C_1\neq 0$  on the boundary and equals to $-D(D-1) k^2/4$ for $C_1= 0$. The non-vanishing components of the Ricci tensor read
\begin{subequations}\label{RMN2}
	\begin{eqnarray}
	    R_{\mu\nu}\!&\!=\!&\!-\frac{D-1}{\Theta}\frac{k^2}{4}g_{\mu\nu}\,,\\
    	R_{rr}\!&\!=\!&\!\Big(\frac{D-3}{\Theta}-D+2 \Big)\frac{(D-1)k^{2}}{4}g_{rr},\\
    	R_{\theta\theta}\!&\!=\!&\!-\frac{(D-1) k^2}{4\Theta}g_{\theta\theta}\,.
\end{eqnarray}
\end{subequations}
If $C_1=0$, the above Ricci tensor reduces to $R_{MN}=-\frac{(D-1)k^2}{4}g_{MN}$, so it describes an $\text{AdS}_{D}$ spacetime. However, if $C_1\neq 0$, the curvature tensor does not describe an asymptotically maximally symmetric spacetime (see the right-hand side of the expression~\eqref{RMN2} in the limit $r\rightarrow\infty$). The behavior of the metric components and the curvature scalar are shown in Fig.~\ref{solution1}. Moreover, the scalar field and the scalar potential are described in Fig.~\ref{phiV1}. As we have mentioned before, we have a constraint between the GB coupling constant and the parameter $k$: $\alpha\geq2/\big[(D-3)k^{2}\big]$. Thus, the first solution could not recover the one in GR by taking the limit $\alpha\rightarrow0$.

\subsection{$H(r)=\text{sech}^{\beta}(kr)$}\label{subsec2}

In this subsection, we assume that the metric component $H$ is
\begin{equation}
	H=\text{sech}^{\beta}(kr),\label{H2}
\end{equation}
where $\beta$ is an arbitrary parameter and $k$ a positive parameter. With the field equation~\eqref{feq3c1}, the solution of the variable $W$ is
\begin{equation}
	W=\text{sech}^{\beta/2}(kr).\label{W2}
\end{equation}
By substituting the above expressions into the field equations~\eqref{feq3c2} and~\eqref{feq3c3}, we obtain the first derivative of the scalar field and the scalar potential:
\begin{eqnarray}
	\phi'&=&k \sqrt{\beta  (D-2)}\sqrt{\chi_{1}(r)}\,\text{sech}(k r) ,\label{phi3}\\
	V&=&\frac{ \beta  (D-2) k^2 }{8}\big[4-\chi_{2}(r)\tanh ^2(k r)\big],\label{V2}
\end{eqnarray}
where
\begin{subequations}
\begin{eqnarray}
	\chi_{1}(r)\!&\!=\!&\!1-\frac{\alpha  \beta ^2 (D-3) k^2}{2} \tanh ^2(k r),\label{chi1}\\
	\chi_{2}(r)\!&\!=\!&\!\beta  (D\!-\!1) (\chi_{1} \!+\!1)\!+\!4 \chi_{1}\!+\!2 \alpha  \beta ^2 (D\!-\!3) k^2.~~~~\label{chi2}
\end{eqnarray}
\end{subequations}
With the expression of $\phi'$~\eqref{phi3} and the definition of $\chi_{1}$~\eqref{chi1}, we find that a real scalar field requires the parameter $\beta$ to satisfy $0\leq\beta\leq1/\sqrt{\alpha(D-3)k^{2}/2}$ for the case $\alpha\neq0$, and $\beta\geq0$ for the case $\alpha=0$. The scalar field could be obtained by integrating the expression~\eqref{phi3}:
\begin{equation}
	\phi=k \sqrt{\beta  (D-2)}\int\sqrt{\chi_{1}}\,\text{sech}(k r) dr\,.\label{phi4}
\end{equation}
Note that the result of the above integration is complicated, so we are not going to show it here. Nevertheless, we find that the expression of the scalar field could be simplified with some specific choices of the parameters $D$, $k$, $\alpha$, and $\beta$. Thus, we give the following two forms of the scalar field as illustrations:
\begin{subequations}
	\begin{eqnarray}
    	\phi_{1}\!&\!=\!&\!2^{3/4} \tan ^{-1}\!\big[\tanh (k r/2)\big],\label{phi5}\\
    	\phi_{2}\!&\!=\!&\!\frac{1}{2^{7/4}}\sqrt{\frac{\text{sech}^2(k r)\!+\!7}{7 \cosh (2 k r)\!+\!9}} \Bigg\{\!\sinh (k r) \sqrt{7 \cosh (2 k r)\!+\!9}\nonumber\\
    	&~&+4 i \cosh (k r) \bigg[E\Big(i k r\Big|\frac{7}{8}\Big)-F\Big(i k r\Big|\frac{7}{8}\Big)\bigg]\Bigg\},\label{phi6}
    \end{eqnarray}
\end{subequations}
where $F\big(ik r|7/8\big)$ and $E\big(i kr|7/8\big)$ are the elliptic integrals of the first and second kinds, respectively. Here, we have set the parameters to $D=4$, $\beta=1/2$, and $\alpha=0$ for~\eqref{phi5} and $\alpha=1/k^{2}$ for~\eqref{phi6}.

Recalling the expressions~\eqref{phi3} and~\eqref{V2}, the energy density of the scalar field is then written as
\begin{eqnarray}
	\rho(r)\!&\!=\!&\!\frac{\beta  (D\!-\!2) k^2 \big[ (\!-\beta \!+\!\beta  D\!+\!8)\chi _1\!+\!C_{4}\big]}{8}\text{sech}^2(k r)\nonumber\\
	&~&-\frac{  C_4 \beta(D\!-\!2) k^2}{8}\!-\!\frac{\beta ^2 (D\!-\!2) (D\!-\!1) k^2}{8}\!-\!\rho_{0}\,,~~~~\label{rho2}
\end{eqnarray}
where the vacuum energy density $\rho_{0}$ is given by the minimum of the scalar potential $V_{0}$ and
\begin{equation}\label{pc4}
	C_{4}=\beta  (D-1) \bigg[1-\frac{\alpha  \beta ^2 (D-3) k^2}{2}\bigg].
\end{equation}
Recalling the expression of the scalar potential~\eqref{V2}, we know that the first and second derivatives of the scalar potential are respectively
\begin{eqnarray}
	V'\!&\!=\!&\!-\frac{(D-2) k^3 \left(C_5 \chi _1\!+\!2 C_4\right)}{2 (D-1)}\tanh (k r) \text{sech}^2(k r),~~~~\label{Vd1c2}\\
	V''\!&\!=\!&\!\frac{\beta  (D-2) k^4}{2} (2 C_6\!+\!\chi _1 \chi _3) \text{sech}^4(k r),\label{Vd2c2}
\end{eqnarray}
with
\begin{subequations}
\begin{eqnarray}
	C_{5}\!&\!=\!&\!-\beta  (D\!-\!1) \big[\beta  (D\!-\!1)\!+\!4\big],\\
	C_{6}\!&\!=\!&\! \beta \bigg[D\!-\!1\! -\! \frac{(D\!-\!3)  \alpha \beta k^2}{2}\bigg]\!+\!7,\\
	\chi_{3}(r)\!&\!=\!&\!\big[\beta  (D\!-\!1)\!+\!2\big] \cosh (2 k r)\!-\!18\!-\!4 \beta  (D\!-\!1).~~~~
\end{eqnarray}
\end{subequations}
The expression~\eqref{Vd1c2} shows two extrema of the scalar potential at $r=0$ and at infinity. By taking into account the expression~\eqref{Vd2c2}, we find that the scalar potential has a maximum at $r=0$ and a minimum $V_{0}$ at infinity. Thus, the vacuum energy density is
\begin{eqnarray}
	\rho_{0}=V_{0}=-\frac{\beta(D-2)k^{2}}{8}\Big[ \beta	 (D-1)+C_4\Big].
\end{eqnarray}
Again, we find that the contribution of the GB term, which is labeled by the GB coupling constant $\alpha$, appears in the vacuum energy density.

To investigate the property of the spacetime with the existence of the cosmic string, we shall calculate the curvature tensor and its products. The curvature scalar is
\begin{eqnarray}\label{R2}
	R(r)&=&-\frac{\beta k^2}{4}\big[D\chi_{4}(r)-2(D-2)\big] \text{sech}^2(k r),
\end{eqnarray}
with
\begin{equation}
	\chi_{4}(r)=\beta  (D-1) \sinh ^2(k r)-2.
\end{equation}
Note that the curvature scalar approximates to $-\beta ^2 (D-1) D k^2/4$ on the boundary. With the expression~\eqref{RMN1} and the metric solution, we show the nonvanishing components of the Ricci tensor as follows:
\begin{subequations}\label{RMN3}
	\begin{eqnarray}
		R_{\mu\nu}&=&-\frac{\beta  k^2}{4}\chi_{4}\,\text{sech}^{2}(k r)g_{\mu\nu}\,,\\
		R_{rr}&=&-\frac{\beta k^2}{4} \big[\chi_{4}-2(D-2)\big]\text{sech}^2(k r)\,g_{rr}\,,~\\
		R_{\theta\theta}&=&-\frac{\beta  k^2}{4} \chi_{4}\,\text{sech}^{2}(k r)\,g_{\theta\theta}\,.
	\end{eqnarray}
\end{subequations}
One could check that the Ricci tensor approximates to
\begin{equation}
	R_{MN}|_{r\rightarrow\infty}=-\frac{\beta ^2 (D-1) k^2}{4}  g_{MN}|_{r\rightarrow\infty}
\end{equation}
on the boundary. Thus, the solution gives an asymptotically $\text{AdS}_{D}$ spacetime.
\begin{figure}[!htb]
\center{
\subfigure[]{\includegraphics[width=3.9cm]{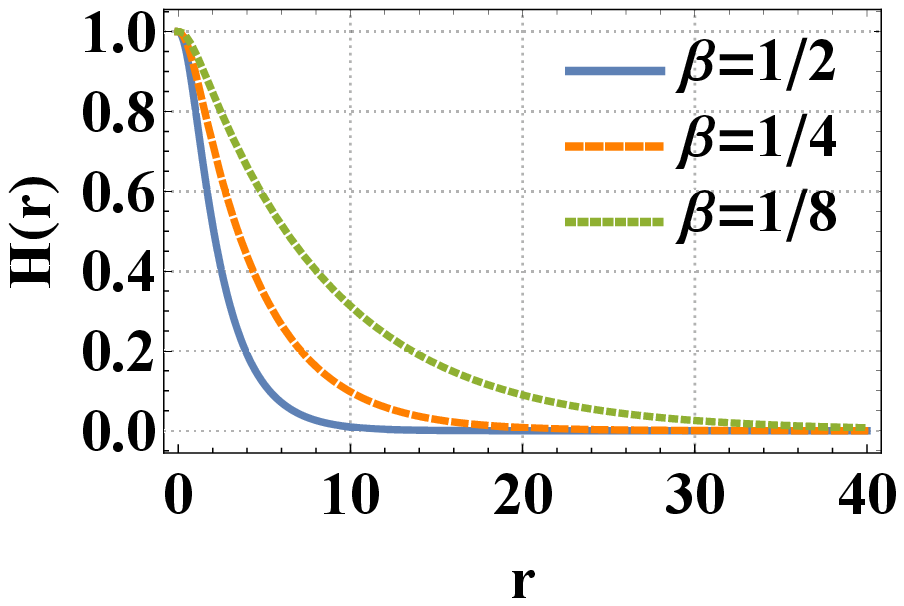}\label{Hc2}}
\quad
\subfigure[]{\includegraphics[width=3.9cm]{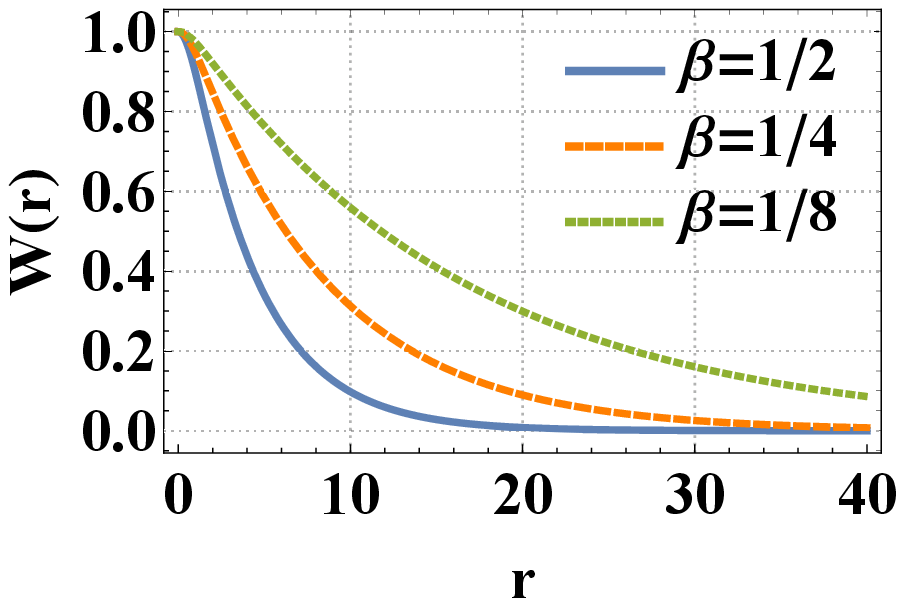}\label{Wc2}}
\quad
\subfigure[]{\includegraphics[width=3.9cm]{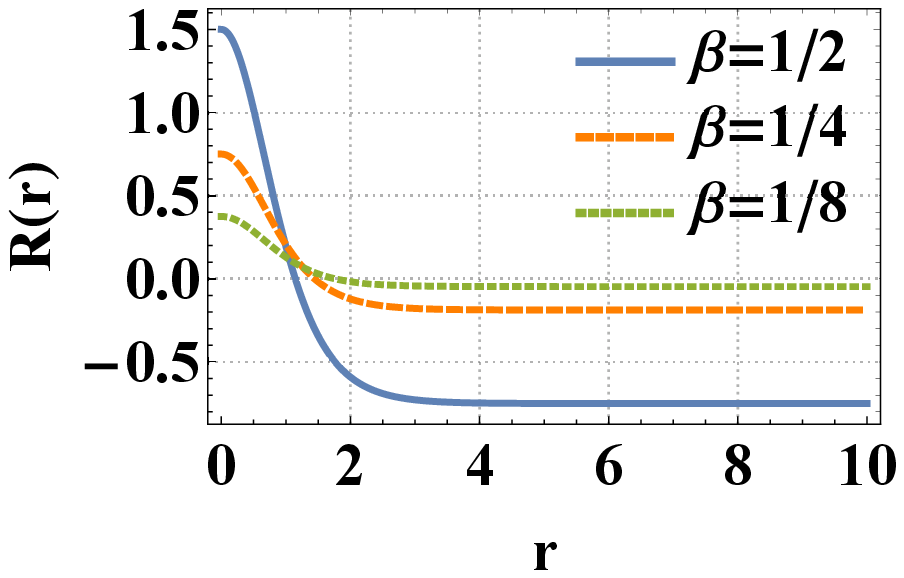}\label{Rc2}}
}
\caption{The metric components~\eqref{H2}-\eqref{W2}, and the curvature scalar~\eqref{R2} in the limit $D\rightarrow4$. (a) The metric component $H(r)$. (b) The solution of $W(r)$. (c) The curvature scalar $R(r)$. The parameter $k$ is fixed to $k=1$.}
\label{solution2}
\end{figure}
The behavior of the metric components and the curvature scalar are plotted in Fig.~\ref{solution2}.
\begin{figure}[!htb]
\center{
\subfigure[]{\includegraphics[width=3.9cm]{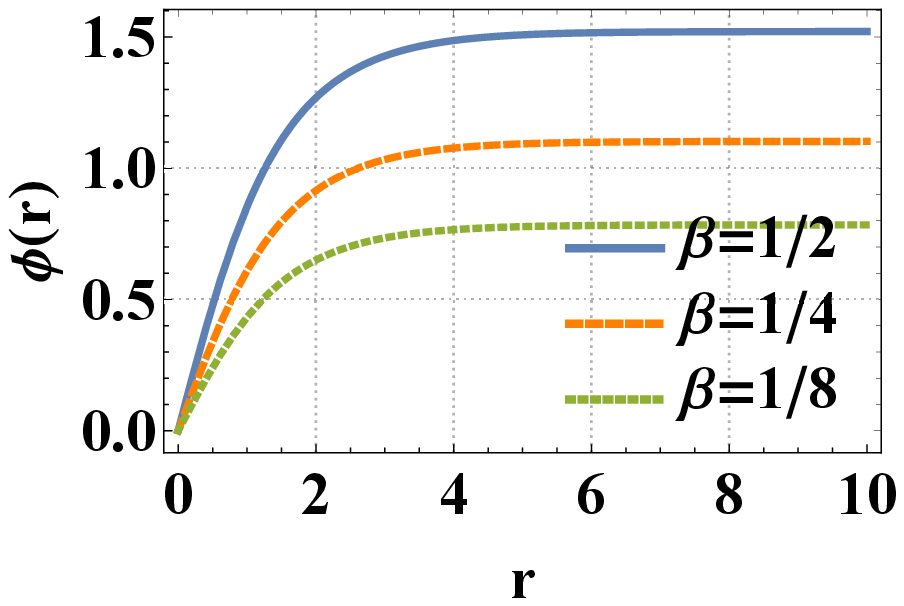}\label{phi1c2}}
\quad
\subfigure[]{\includegraphics[width=3.9cm]{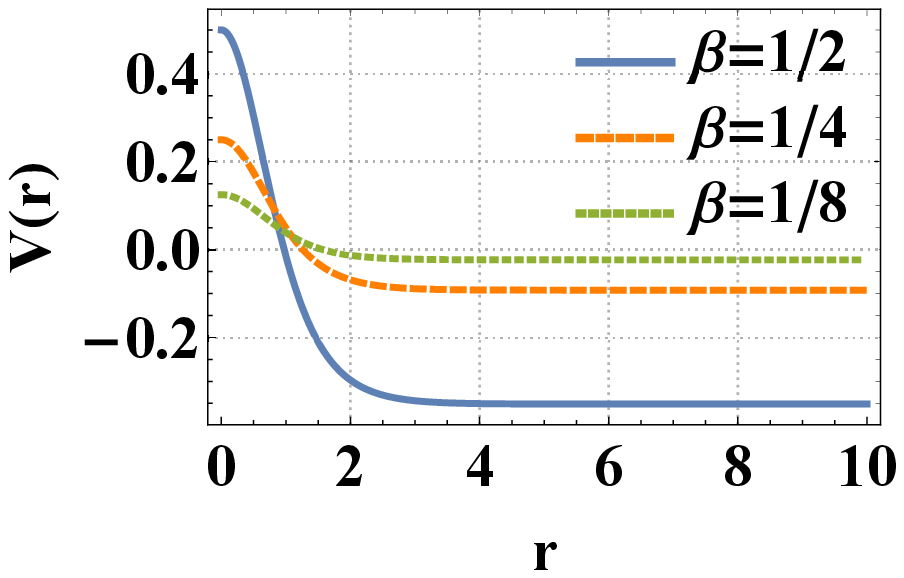}\label{V1c2}}
\quad
\subfigure[]{\includegraphics[width=3.9cm]{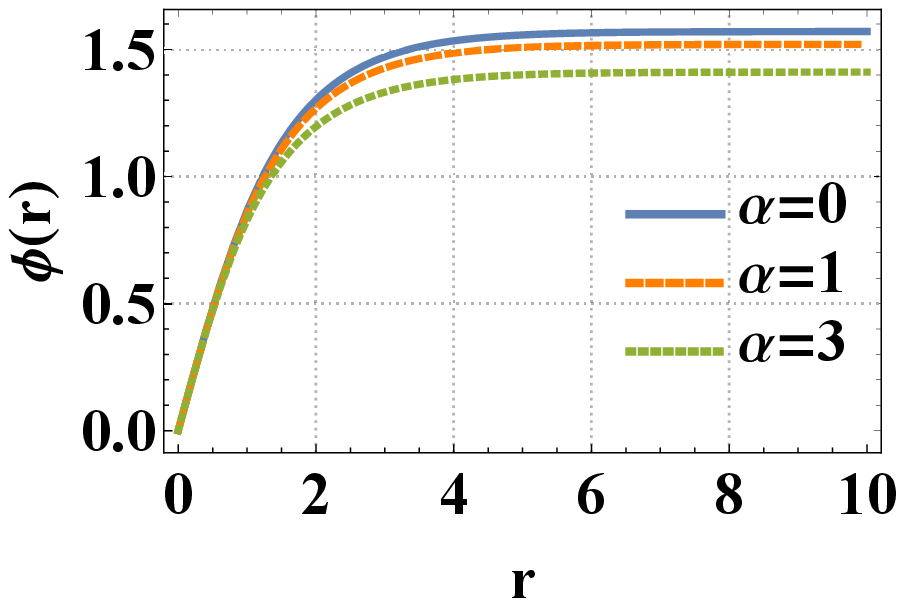}\label{phi2c2}}
\quad
\subfigure[]{\includegraphics[width=3.9cm]{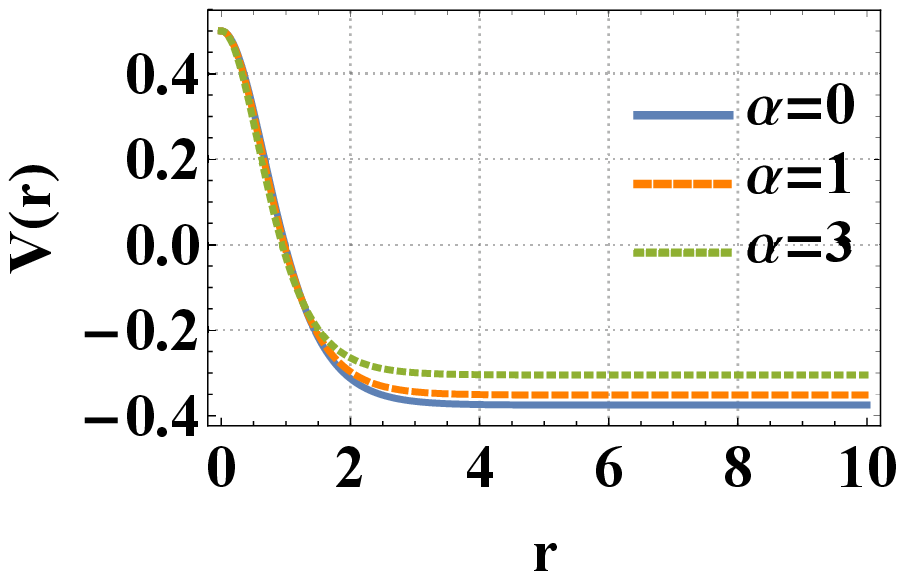}\label{V2c2}}
}
\caption{The scalar field (\ref{phi4}) and the scalar potential (\ref{V2}) at the limit $D\rightarrow4$. (a) The scalar field $\phi(r)$ and (b) the scalar potential for different values of the parameter $\beta$ and $\alpha=1$. (c) The scalar field and (d) the scalar potential for different values of the GB coupling constant and $\beta=1/2$. The parameter $k$ is fixed to $k=1$.}
\label{phiv2}
\end{figure}
Obviously, all these quantities are smooth in the $r$ direction. Thus our solution obtained in this section gives a regular spacetime. Curves with different values of the parameter $\beta$ are plotted aiming to denote effects from the scalar field. Comparing Fig.~\ref{solution2} with Figs.~\ref{phi1c2} and~\ref{V1c2}, we directly show how the scalar field gives contribution to the spacetime geometry. We then fix the parameter $\beta$ and vary the coupling constant to reveal that the contributions of the GB term in field equations behave like an effective matter content. Figs.~\ref{phi2c2} and~\ref{V2c2} indicate that the same spacetime geometry might result from different scalar field solutions with different values of the GB coupling constant. As one can check, the regularity of the spacetime geometry around the scalar field also holds when the EGB theory recovers GR (e.g. $\alpha\rightarrow0$).
\begin{figure}[!htb]
\center{
\subfigure[]{\includegraphics[width=3.9cm]{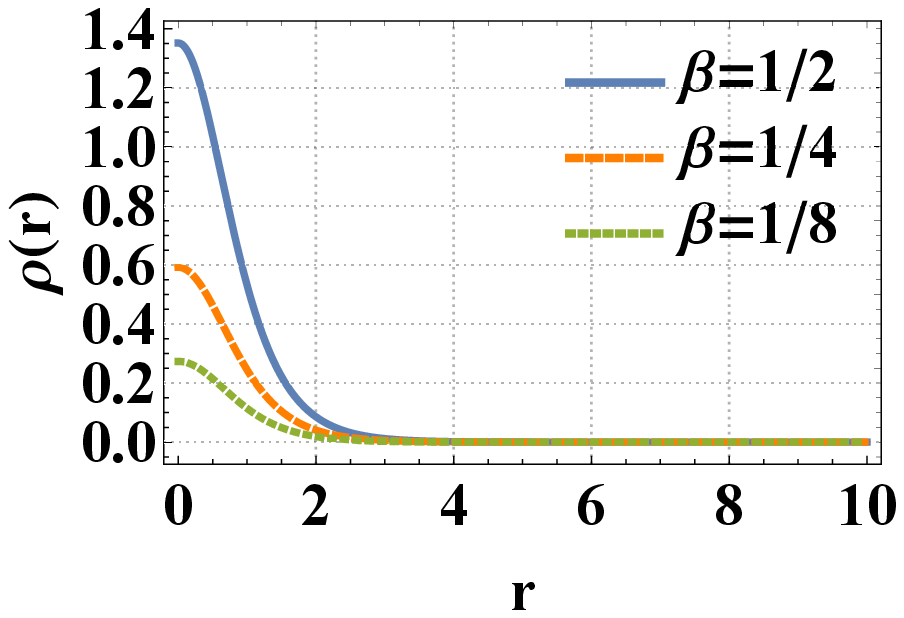}\label{rho1c2}}
\quad
\subfigure[]{\includegraphics[width=3.9cm]{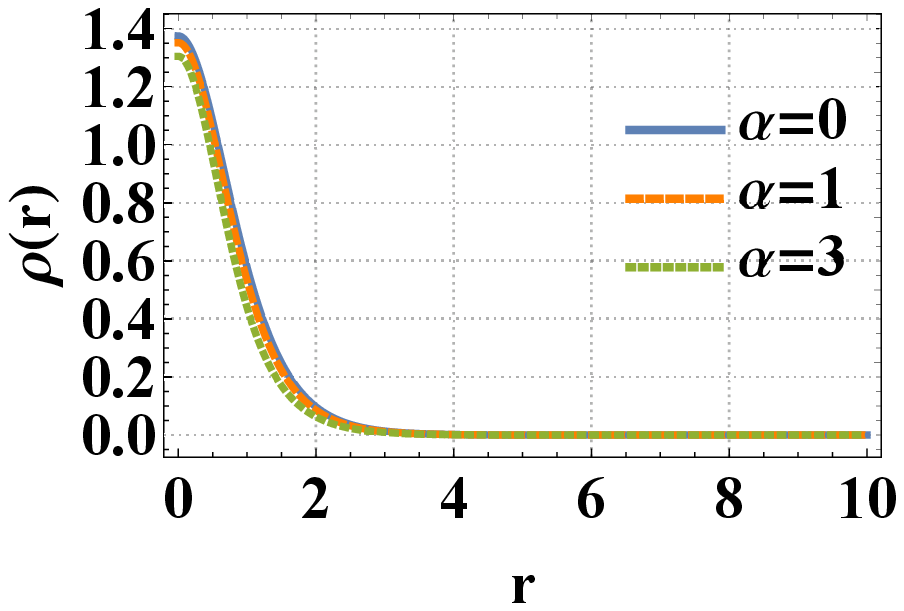}\label{rho2c2}}
}
\caption{The energy density~\eqref{rho2} of the scalar field in the $D\rightarrow4$ limit. (a) The energy density for different values of the parameter $\beta$ and  $\alpha=1$. (b) The energy density for different values of the GB coupling constant and $\beta=1/2$. For both cases, we set the parameter $k$ to unit.}
\label{rhoc2}
\end{figure}

The energy density of the scalar field is shown in Fig.~\ref{rhoc2}, from which one can find a smooth distribution of the energy density along the coordinate $r$. It indicates that the cosmic string has a width $\sim 1/k$ in $r$ direction and locates around the $z$ axis. Besides, Fig.~\ref{rho1c2} also shows an energy density increasing with the parameter $\beta$. It is natural to understand the dependency since a vacuum results in a Minkowski spacetime (e.g. $\beta\rightarrow0$ when $\rho(0)\rightarrow0$). Fig.~\ref{rho2c2} also supports that the spacetime geometry is affected by both the matter field and the GB term. All these results indicate that the EGB theory deviates from GR in the limit $D\rightarrow4$ with the redefinition $\alpha\rightarrow\alpha/(D-4)$.

Up to now, we have obtained the solution of the metric components, \eqref{H2} and~\eqref{W2}, which describe the geometry around the cosmic string. We have also calculated the energy density~\eqref{rho2} of the cosmic string. Another important property of the cosmic string is its mass per unit length along its axis. Now we analyze how the mass density of the cosmic string changes with respect to the GB coupling constant and the spacetime dimension. It may help us to understand the effects from the GB term and higher dimensions on the energy scale of the phase transition that arises cosmic strings. Since the metric has an axial symmetry, we use a piece of the cosmic string, which has a coordinate distance $z_{2}-z_{1}$ from one end of the string to the other, to define its mass density:
\begin{eqnarray}
	 m(r)\!&\!=\!&\!\int^{2\pi}_{0}\!\int^{\infty}_{0}\!\int^{z_{2}}_{z_{1}}\!\rho(r')W(r')\sqrt{H(r')}\mathcal{V}(r')\,d\theta dr'dz\nonumber\\
	 \!&\!~\!&\!\times\ell(r)^{-1},
\end{eqnarray}
where $\mathcal{V}$ is the volume of the rest $(D-4)$-dimensional space defined as
\begin{equation}
	\mathcal{V}(r')=\int_{\mathcal{V}} H(r')^{\frac{D-4}{2}} d^{D-4}x,
\end{equation}
and the proper length $\ell(r)$ of the string is measured by
\begin{equation}
	\ell(r)=\int^{z_{2}}_{z_{1}}\sqrt{H(r)}\,dz,
\end{equation}
where $r$ is the coordinate distance from the observer to the axis $z$. As we have mentioned before the energy density~\eqref{rho2} of the string indicates that the string has a width (see Fig.~\ref{rhoc2}). Thus, for the sake of simplification, we assume that $\ell(0)$ is the proper length of the piece of the string we choose. Accordingly, the mass per unit length of the cosmic string is
\begin{eqnarray}
	 m_{0}\!&=&\!m(0)\!=\!\int^{2\pi}_{0}\!\int^{\infty}_{0}\!\rho(r')W(r')\sqrt{\!\frac{H(r')}{H(0)}}\mathcal{V}(r')\,d\theta dr'.\nonumber\\
	 ~&~&\label{m02}
\end{eqnarray}
In Table~\ref{m0T1c2}, we list the values of the string mass density for different sets of parameters. It is easy to find that the GB term contributes to the gravity in the $D\rightarrow4$ limit, and the string mass density decreases with it. An important feature is that the number of dimension dilutes the contribution of string mass on the spacetime geometry, by comparing the profile with different dimensions in Table~\ref{m0T1c2}. As one can see, for the same solution of the metric components $H$ and $W$ (with fixed $\beta$), a larger mass density of the comic string is required when the spacetime dimension increases. Thus, for the same mass density of the cosmic string, when the number of the spacetime dimension increases, the parameter $\beta$ decreases. In other words, the gravitational radius of the cosmic string decreases when the spacetime dimension increases (see Fig.~\ref{Rc2}).
\begin{table*}
	\begin{tabular}{cccc}
		\hline\hline
		~Dimension $D$~~&~GB Coupling constant $\alpha$~~&~$\beta$~~&~Mass density $m_{0}$~~\\
		\hline
		 \multirow{5}{*}{~$D\rightarrow4$~~}&~0~~&~1/2~~&~7.55~~\\
		 &1&1/2&7.19\\
		 &3&1/2&6.47\\
		 &1&1/4&3.43\\
		 &1&1/8&1.65\\
		\hline
		~$D=5$~~&~1~~&~1/2~~&~11.11~~\\
		\hline
		~$D=6$~~&~1~~&~1/2~~&~~15.02~~
		\\ \hline
	\end{tabular}
	\caption{The mass density~\eqref{m02} of the cosmic string for different values of the GB coupling constant $\alpha$  and string thickness. We choose $\alpha=0$, $\alpha=1$, and $\alpha=3$ in order to reveal the contributions of the GB term on the spacetime geometry. The mass densities with different values of the parameter $\beta$ reflect the contribution of the string mass to the spacetime geometry. The parameter $D$ is respectively set to $D\rightarrow4$, $D=5$, and $D=6$ to show the variation of the string mass density in different dimensions. In each set, the parameter $k$ is fixed to unit.}\label{m0T1c2}
\end{table*}

\subsection{$H(r)=(r+\sqrt{D-3}\sqrt{2\alpha}\,\sigma/2)^{\sigma}$}\label{subsec3}

In this subsection, we give another set of the comic string solution. As usual, we will first give an assumption of the metric component $H$. As was pointed out in Refs.~\cite{Hennigar3,Olasagasti1,Gherghetta1,Oda1,Dzhunushaliev1,Gregory1,Momeni1,Sharif1}, for a cosmic string spacetime, the metric component could have a form of $\sigma$th power of the coordinate $r$ when GR is recovered. Thus, we assume
\begin{equation}\label{H3}
	H=(r+\sqrt{D-3}\sqrt{2\alpha}\,\sigma/2)^{\sigma},
\end{equation}
where $\sigma$ is a positive parameter. By taking into account the field equation~\eqref{feq3c1}, we can obtain the expression of the metric component $W$:
\begin{equation}\label{W3}
	W=(r+\sqrt{D-3}\sqrt{2\alpha}\,\sigma/2)^{\sigma/2}.
\end{equation}
Note that both metric components~\eqref{H3} and~\eqref{W3} contain the GB coupling constant, therefore we could obtain the corresponding metric components in GR by taking the $\alpha\rightarrow0$ limit. Recalling the other nonvanishing components of the field equations, Eqs.~\eqref{feq3c2} and~\eqref{feq3c3}, we then get the following expressions of the first derivative of the scalar field and the scalar potential
\begin{eqnarray}
	 \phi'\!&\!=\!&\!\sqrt{\sigma}(2C_{7})^{1/4}\sqrt{\varsigma_{2}}\,\varsigma_{3}\,\varsigma_{1}^{-2},\label{phi7} \\
	V\!&\!=\!&\!\frac{(D\!-\!1) \sigma\!-\!4}{16(D\!-\!2)}\frac{C_{7}\,\sigma}{\varsigma_{1}^{4}}\!-\!\frac{  (D\!-\!1) \sigma \!-\!2}{4}\frac{(D\!-\!2) \sigma}{\varsigma_{1}^{2}},~~~~\label{V3}
\end{eqnarray}
where
\begin{subequations}
\begin{eqnarray}
    \varsigma_{1}(r)&=&r+\sqrt{D-3}\sqrt{2\alpha}\,\sigma/2,\\
	\varsigma_{2}(r)&=&r+ \sqrt{D-3}\sqrt{2\alpha } \,  \sigma ,\\
	\varsigma_{3}(r)&=&\sqrt{D-2}(2C_{7})^{-1/4}\sqrt{r},\\
	C_{7}&=&\alpha(D-3)(D-2)^{2}\sigma^{2}.
\end{eqnarray}
\end{subequations}
By integrating the first derivative of the scalar field~\eqref{phi7}, we find the expression of the scalar field
\begin{equation}
	\phi=\Big(\frac{C_{7}}{\alpha}\Big)^{1/4} \big(2  \sinh ^{-1}(\varsigma_{3} )-\sqrt{r} \sqrt{\varsigma_2}\,\varsigma_{1}^{-1}\big).\label{phi8}
\end{equation}
By taking into account the expressions~\eqref{phi7} and~\eqref{V3}, we then obtain the energy density of the scalar field
\begin{eqnarray}
	\rho(r)&=&\frac{(D-1) \sigma -4}{16 (D-2)}\frac{C_{7}\,\sigma}{\varsigma_{1}^{4}}+
	 \sqrt{\frac{C_{7}}{2}}\frac{\sigma\,\varsigma_{2}\,\varsigma_{3}^{2}}{\varsigma_{1}^{4}}\nonumber\\
	&~&-\frac{(D-2) \sigma  \big[(D-1) \sigma -2\big]}{4\varsigma_{1}^{2}}\label{rho3}
\end{eqnarray}
with a vanishing vacuum energy density $\rho_{0}=0$. Thus, this set of solution gives a vanishing effective cosmological constant. And no contribution from the GB term appears on the vacuum energy density of the scalar field.

We now investigate the property of the spacetime with the existence of the cosmic string. Taking the trace of the Ricci tensor~\eqref{RMN1}, we get the curvature scalar as follows:
\begin{equation}\label{R3}
	R(r)=-\frac{(D-1)  (D \sigma -4)\sigma }{4 \varsigma_{1}^2},
\end{equation}
which vanishes on the boundary. The nonvanishing components of the Ricci tensor are
\begin{subequations}
\begin{eqnarray}
	R_{\mu\nu}&=&\frac{2-(D-1) \sigma }{4\varsigma_1^{2}} \sigma \,g_{\mu\nu}\,,\\
	R_{rr}&=&-\frac{(D-1) (\sigma -2) \sigma }{4 \varsigma_{1}^2}g_{rr}\,,\\
	R_{\theta\theta}&=&\frac{2-(D-1) \sigma }{4 \varsigma_{1}^{2 }}\sigma\, g_{\theta\theta}\,.
\end{eqnarray}
\end{subequations}
One could check that the prefactors of the metric components $g_{\mu\nu}$, $g_{rr}$, and $g_{\theta\theta}$ vanish on the boundary. Here and after, we will constrain the parameter $\sigma$ to $\sigma<2/(D-2)$, since, as we will discuss later, the mass density of the cosmic string diverges when the parameter $\sigma$ violates this constraint. Then, by taking into account the value of the metric components in the limit $r\rightarrow\infty$, we find that each component of the Ricci tensor vanishes on the boundary. The nonvanishing components of the Riemann curvature tensor~\eqref{RKLMN1} are calculated as
\begin{subequations}
\begin{eqnarray}
	R^{\theta}_{~r\theta r}&=&-\frac{(\sigma -2) \sigma }{4 \varsigma_1^2},\\
	R^{\theta}_{~\mu\theta\nu}&=&-\frac{\sigma ^2}{4 \varsigma_1^{2-\sigma}} \eta_{\mu\nu}\,,\\
	R^{r}_{~\theta\theta r}&=&\frac{ (\sigma -2) \sigma }{4\varsigma_1^{2-\sigma }},\\
	R^{r}_{~\mu r\nu}&=&-\frac{ (\sigma -2) \sigma }{4\varsigma_1^{2-\sigma }}\eta_{\mu\nu}\,,\\
	R^{\mu}_{~\theta\theta\nu}&=&\frac{\sigma ^2}{4\varsigma_1^{2-\sigma}} \delta^{\mu}_{~\nu}\,,\\
	R^{\gamma}_{~\mu\nu\lambda}&=&\frac{\sigma ^2}{2\varsigma_1^{2-\sigma}}\eta_{\mu[\nu}\delta^{\gamma}_{~\lambda]}\,.
\end{eqnarray}
\end{subequations}
Again, with the constraint mentioned above, each component of the curvature tensor vanishes on the boundary. Therefore, the solution obtained in this section describes an asymptotically Minkowski spacetime.

\begin{figure*}[!htb]
\center{
\subfigure[]{\includegraphics[width=3.9cm]{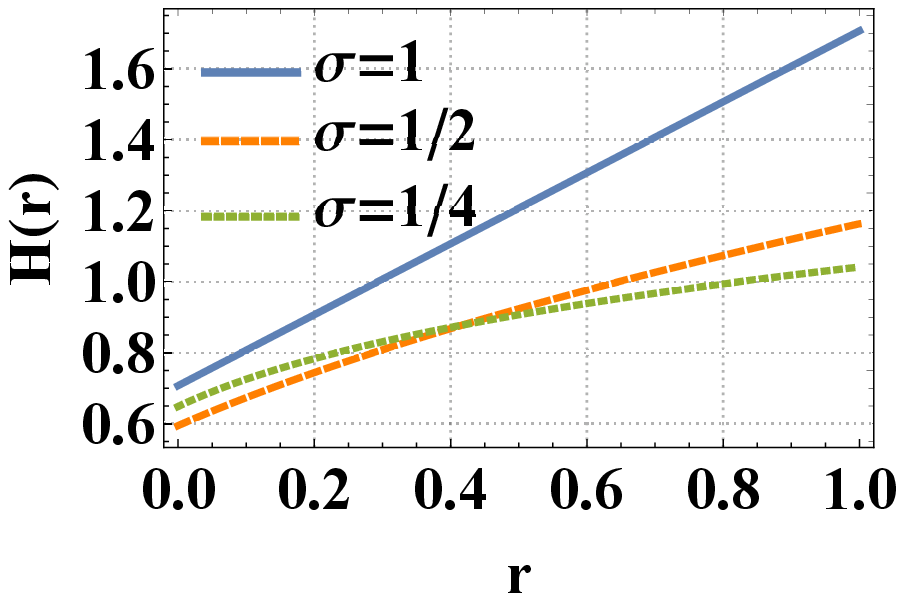}\label{H1c3}}
\quad
\subfigure[]{\includegraphics[width=3.9cm]{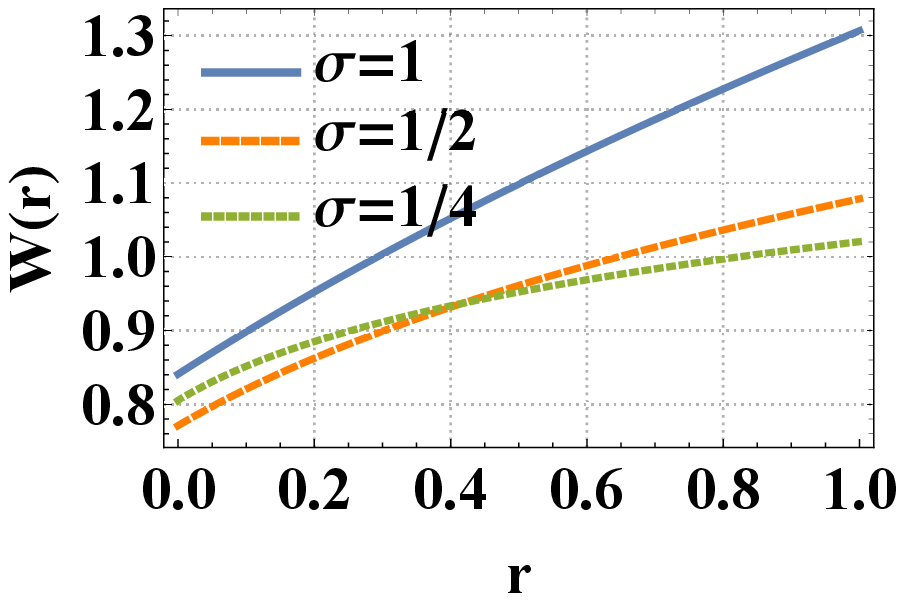}\label{W1c3}}
\quad
\subfigure[]{\includegraphics[width=3.9cm]{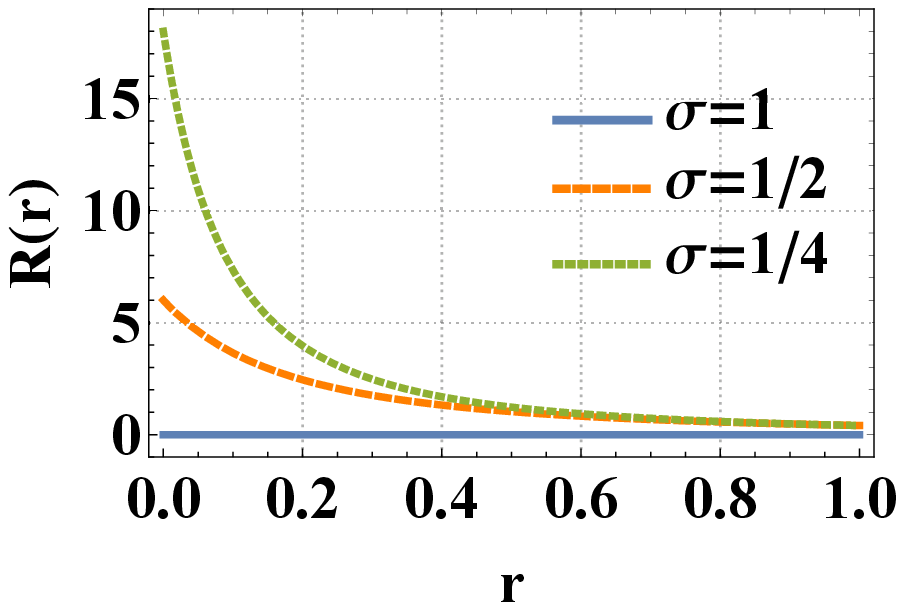}\label{R1c3}}\\
\subfigure[]{\includegraphics[width=3.9cm]{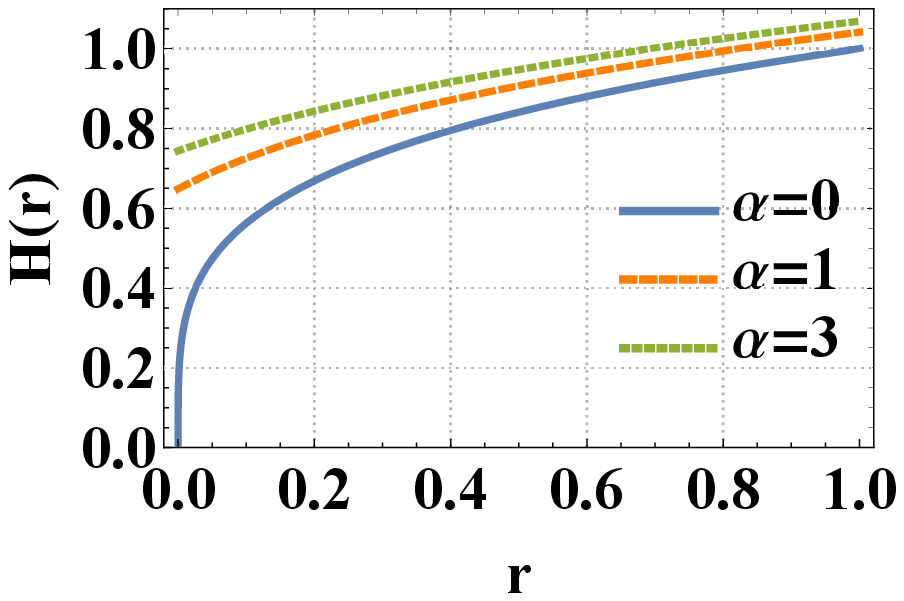}\label{H2c3}}
\quad
\subfigure[]{\includegraphics[width=3.9cm]{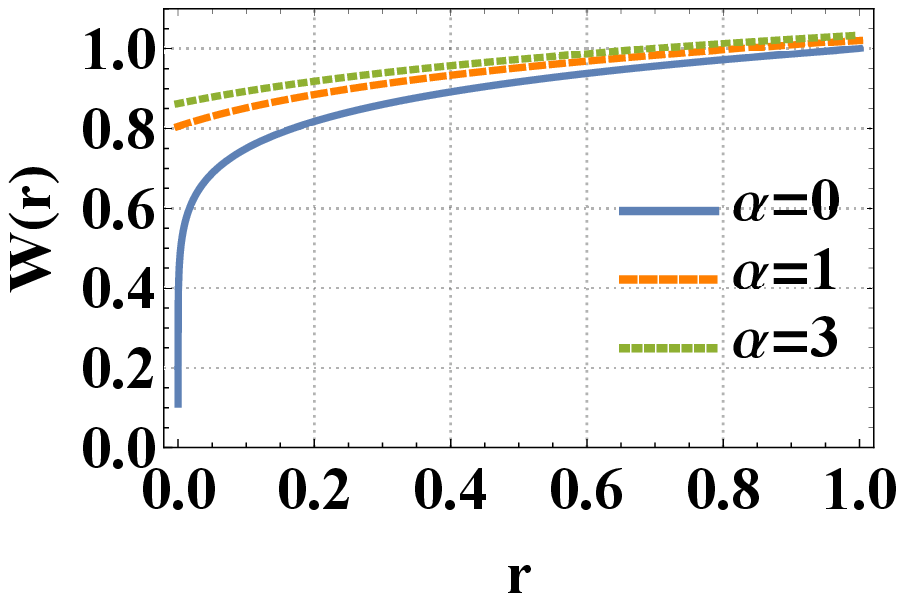}\label{W2c3}}
\quad
\subfigure[]{\includegraphics[width=3.9cm]{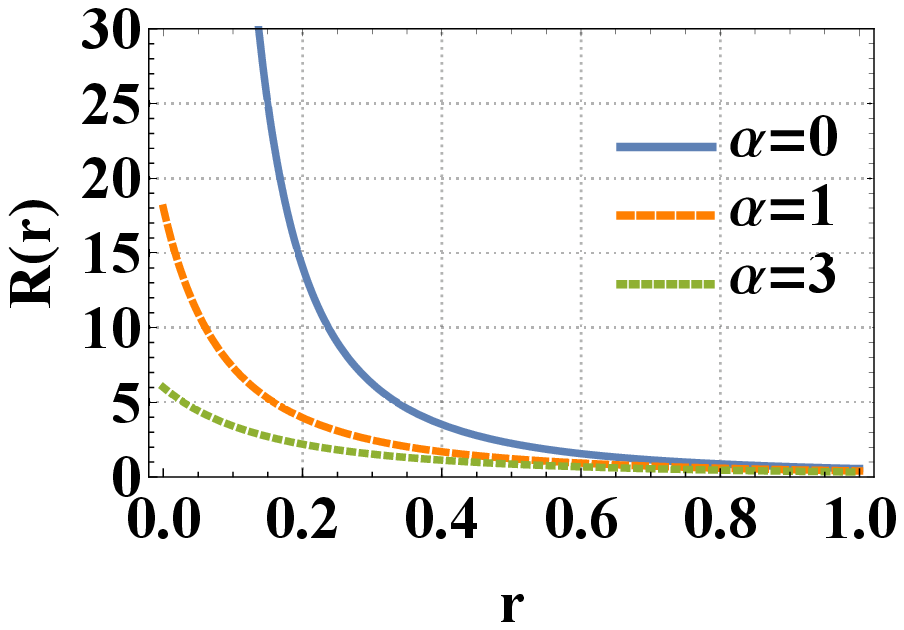}\label{R2c3}}
}
\caption{The metric components~\eqref{H3}-\eqref{W3} and the curvature scalar~\eqref{R3}. (a) The metric component $H(r)$, (b) the solution of $W(r)$, and (c) the curvature scalar $R(r)$ with $\alpha=1$. (d) The metric component $H(r)$, (e) the solution of $W(r)$, and (f) the curvature scalar $R(r)$ with $\sigma=1/4$. The sketch describes the spacetime geometry in the $D\rightarrow4$ limit.}
\label{solution3}
\end{figure*}

From the metric components~\eqref{H3} and~\eqref{W3}, it is clear that the parameter $\sigma$, the spacetime dimension $D$, and the GB coupling constant affect the spacetime geometry. We then take the $D\rightarrow4$ limit and fix the coupling constant to show different spacetime geometries with respect to the parameter $\sigma$ (see Figs.~\ref{H1c3},~\ref{W1c3}, and~\ref{R1c3}). The contributions from the GB coupling constant are also shown in the diagram by fixing the parameter $\sigma$ and taking the limit $D\rightarrow4$. Obviously, curves with a vanishing GB coupling constant correspond to the case of GR. Note that our solution shows that when the GB coupling constant vanishes, a singularity appears at $r=0$ (see blue solid curves in Figs.~\ref{H2c3},~\ref{W2c3}, and~\ref{R2c3}). However, for the case of a nonvanishing GB coupling constant, the contributions of the GB term on the field equations remove such singularity (see also  Figs.~\ref{H2c3},~\ref{W2c3}, and~\ref{R2c3}). These conclusions could be extended into a higher-dimensional version by varying the value of the spacetime dimension.
\begin{figure}[!htb]
\center{
\subfigure[]{\includegraphics[width=3.9cm]{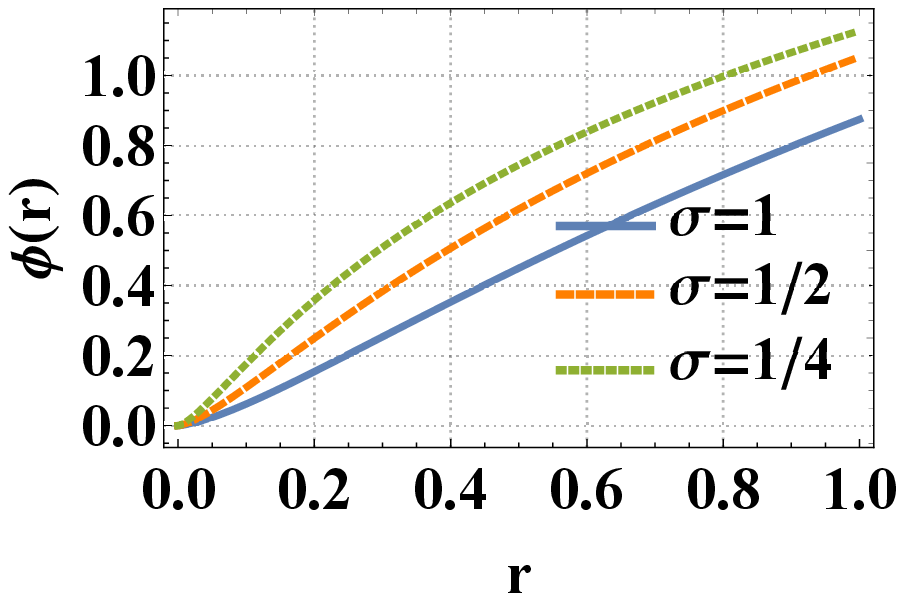}\label{phi1c3}}
\quad
\subfigure[]{\includegraphics[width=3.9cm]{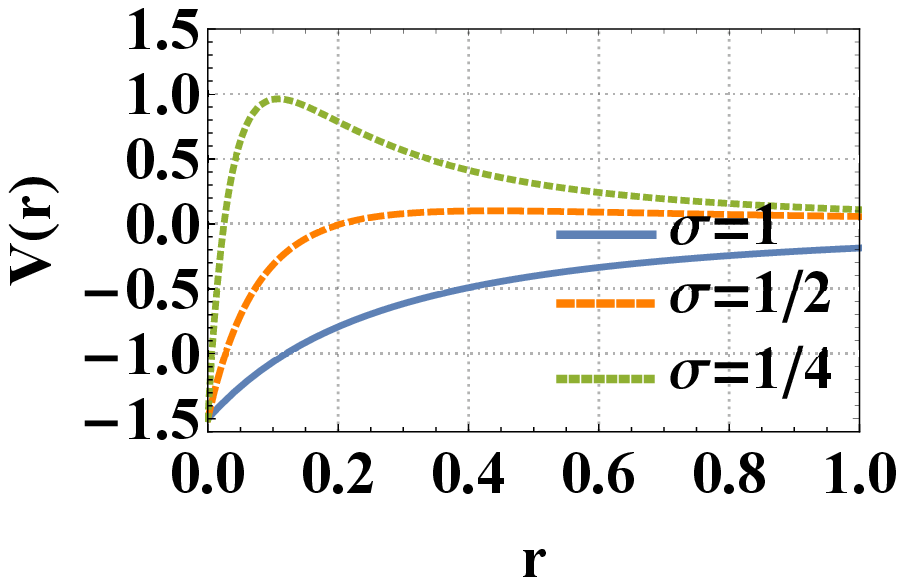}\label{V1c3}}
\quad
\subfigure[]{\includegraphics[width=3.9cm]{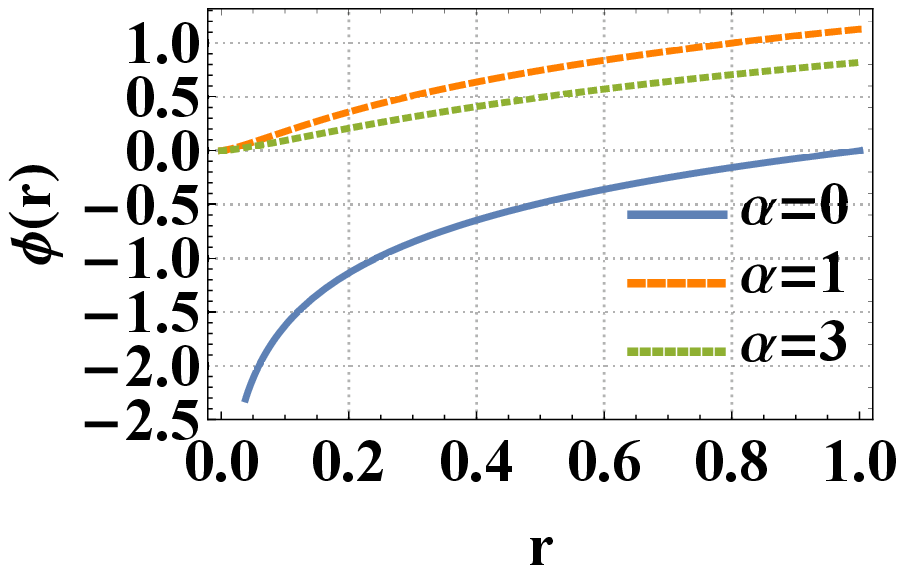}\label{phi2c3}}
\quad
\subfigure[]{\includegraphics[width=3.9cm]{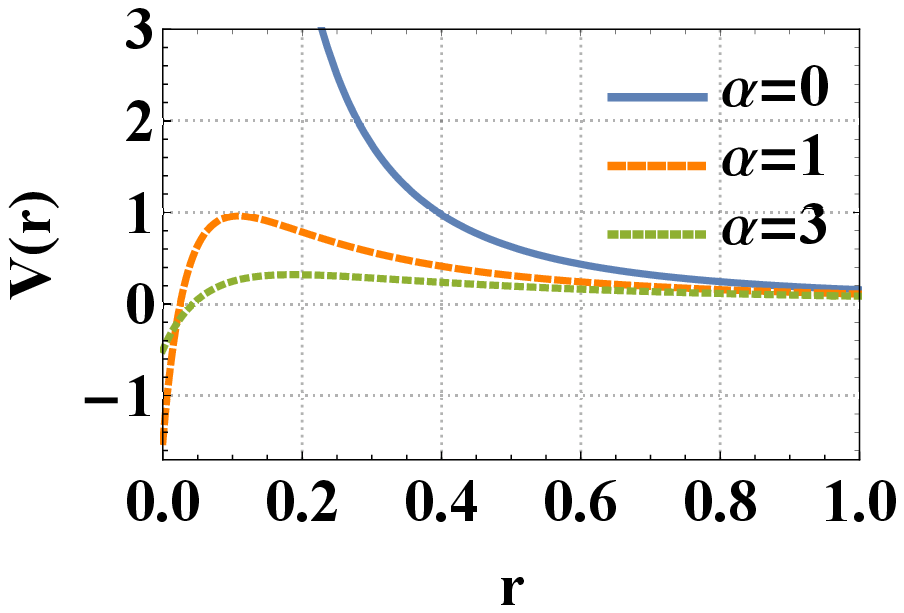}\label{V2c3}}
}
\caption{The scalar field~\eqref{phi8} and the scalar potential~\eqref{V3} at the $D\rightarrow4$ limit. (a) The scalar field $\phi(r)$ and (b) the scalar potential $V(r)$ for different values of the parameter $\sigma$ with $\alpha=1$. (c) The scalar field $\phi(r)$ and (d) the scalar potential $V(r)$ for different values of the GB coupling constant with $\sigma=1/4$.}
\label{phiV3}
\end{figure}

The behaviors of the scalar field and the scalar potential are plotted in Fig.~\ref{phiV3}. From Fig.~\ref{V1c3}, one finds that a potential barrier appears when the parameter $\sigma$ decreases. Such barrier also arises when the GB coupling constant decreases (see Fig.~\ref{V2c3}). Note that Fig.~\ref{V2c3} shows a divergence of the scalar potential at $z$ axis when the GB term vanishes. While the scalar field becomes finite at $r=0$ when the GB term is switched on.
\begin{figure}[!htb]
\center{
\subfigure[]{\includegraphics[width=3.9cm]{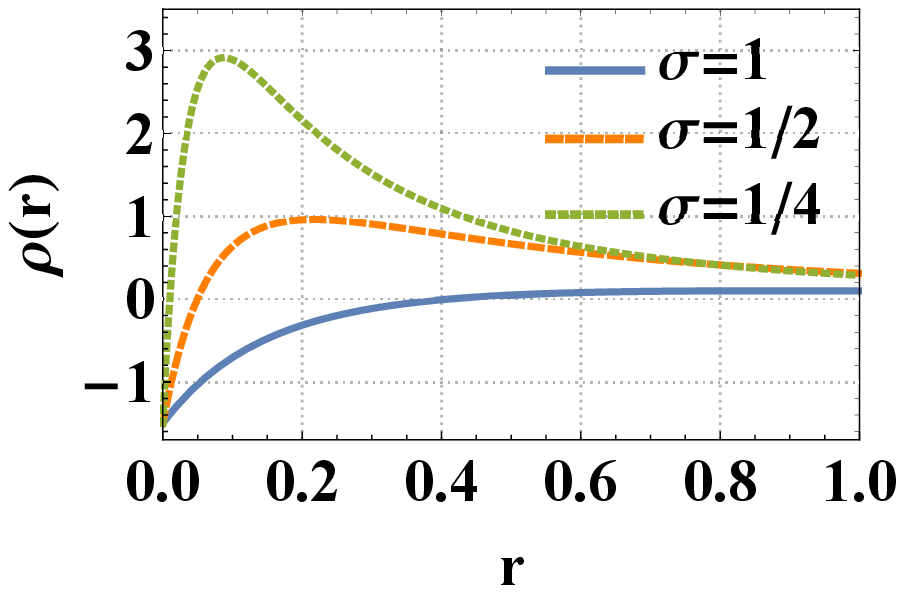}\label{rho1c3}}
\quad
\subfigure[]{\includegraphics[width=3.9cm]{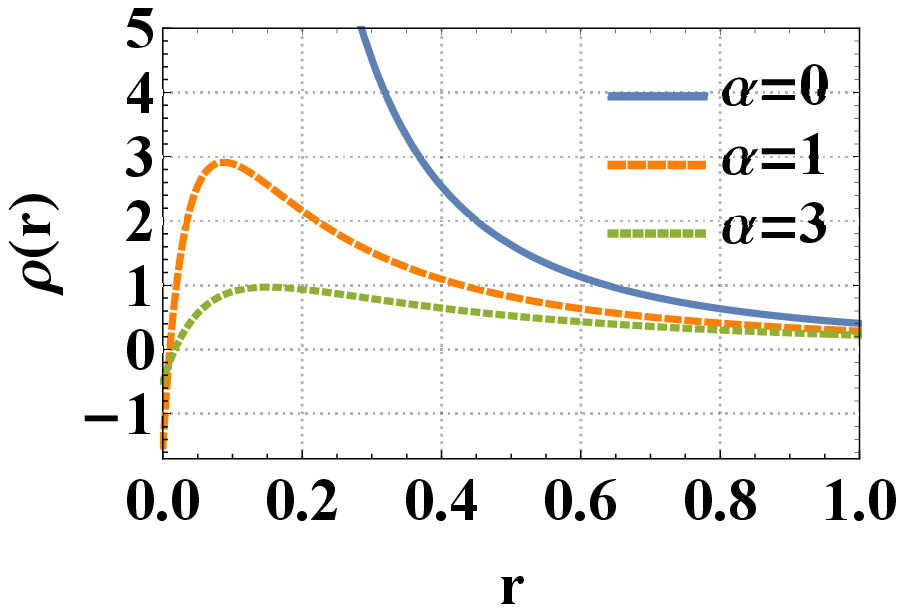}\label{rho2c3}}
}
\caption{The string energy density~\eqref{rho3} at the limit $D\rightarrow4$.  (a) The string energy density for different spacetime geometries with $\alpha=1$. (b) The string energy density for different GB coupling constants with $\sigma=1/4$.}
\label{rhoc3}
\end{figure}

In Fig.~\ref{rhoc3}, we sketch the energy density of the cosmic string in four-dimensional spacetime. For all the cases we choose, the energy density converges on the boundary. While it diverges at origin when the GB term vanishes. Thus, the  energy density of the scalar field is convergent only when the GB coupling constant is nonvanishing.  Figure~\ref{rhoc3} also reveals that the cosmic string obtained in this section has a width and locates near the $z$ axis. Our results support that the GB term gives nontrivial effects on the cosmic string obtained in this section in the $D\rightarrow4$ limit by applying the redefinition $\alpha\rightarrow\alpha/(D-4)$. In the following, we will show how the mass of the cosmic string obtained in this section affects the surrounding spacetime geometry, and how the gravitational radius of the cosmic string decreases with the increase of the spacetime dimension. The nontrivial effect of the GB term as an effective matter content is also shown.

\begin{figure}[!htb]
\center{
\subfigure[]{\includegraphics[width=3.9cm]{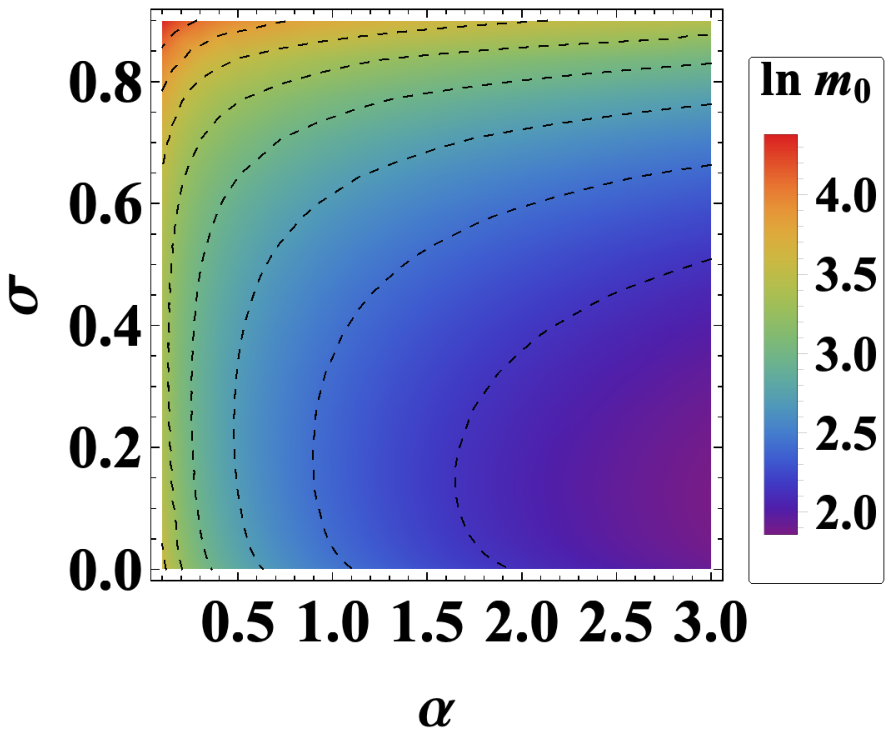}\label{m03c3}}
\quad
\subfigure[]{\includegraphics[width=3.9cm]{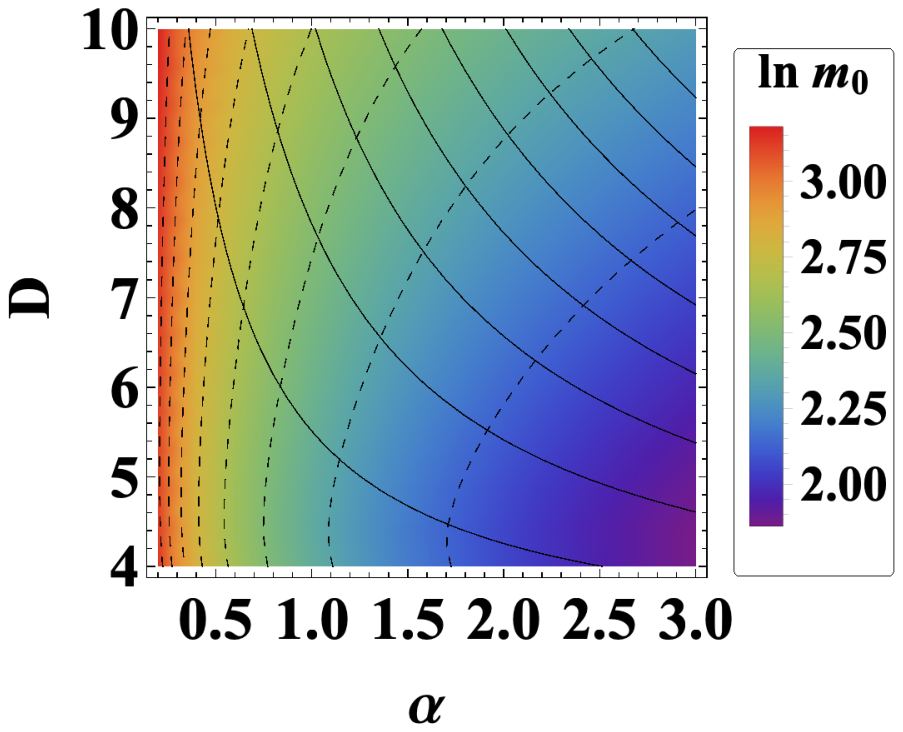}\label{m04c3}}
}
\caption{Plots of the string mass density~\eqref{m03}. (a) The String mass density $m_{0}$ in the parameter space ($\alpha$, $\sigma$) in the limit $D\rightarrow4$. (b) The string mass density $m_{0}$ in the parameter space ($\alpha$, $D$) with $\sigma=1/10$. Each dashed curve has the same value of mass density. And each solid curve corresponds to the same metric solution.}
\label{m0c3}
\end{figure}

Recalling the definition of the mass density of a cosmic string~\eqref{m02}, the string mass density is given as follows:
\begin{equation}
	m_{0}=\frac{ C_{8}C_{9}\,\pi  }{\big(\sqrt{\alpha/2} \sqrt{D-3}\,  \sigma \big)^{1- (D-3) \sigma /2}},\label{m03}
\end{equation}
where
\begin{subequations}
	\begin{eqnarray}
		C_{8}&=&(D-2) \sigma,\\
		C_{9}&=&\frac{(C_8-10) (D-1) \sigma +32}{ (C_8-6) (C_8-2)}.
	\end{eqnarray}
\end{subequations}
Note that we have used the condition $\sigma<2/(D-2)$ to avoid the divergence of the string mass density. Obviously, the string mass density is a function of the parameter $\sigma$, the GB coupling constant $\alpha$, and the spacetime dimension $D$. We further give the parametric diagram of the string mass density in Fig.~\ref{m0c3}, from which one finds that the mass density decreases with the increasing of $\sigma$ or $\alpha$. Each dashed curve in Fig.~\ref{m03c3} labels the cosmic strings with the same mass density. It then indicates that the GB term contributes to the spacetime geometry as an effective matter content. The nontrivial effect from the GB term on the spacetime geometry is also shown through the dashed curves in Fig.~\ref{m04c3}. Therein we use each of them to denote a series of solutions with the same string mass density. In Fig.~\ref{m04c3}, each solid curve has the same value of $\sqrt{D-3}\sqrt{\alpha/2}\,\sigma$. Thus each of them corresponds to the same expression of the metric components (see expressions~\eqref{H3} and~\eqref{W3}). These solid curves indicate that, for the same metric solution, the string mass density increases with the number of the spacetime dimension. In other words, when the spacetime dimension increases, the gravitational radius decreases for a given string mass.

\subsection{Equivalence of the two actions for 4D cosmic string solutions}\label{subsec4}

In this subsection, we will give a short calculation to show that the three sets of solutions obtained in this section are indeed the solutions in the regularized 4D EGB theory. Our result also supports our proof given in the last section.

By taking into account our three sets of solutions, the gravitational field equations, Eqs.~\eqref{r4DEGBgfq1},~\eqref{r4DEGBgfq2}, and~\eqref{r4DEGBgfq3}, can be rewritten as
\begin{subequations}\label{r4EGBgfeq2}
\begin{eqnarray}
    \text{Case A:~}0\!&\!=\!&\!(k\!-\!2 \varphi)\big[(k\!-\!2 \varphi)^2 (3 k\!+\!2 \varphi )\!-\!32 (k\!+\!\varphi ) \varphi '\big],\nonumber\\
    &&\\
	0\!&\!=\!&\!(k\!-\!2 \varphi)\big[C_1 e^{\frac{3}{2} k r} (3 k\!+\!2 \varphi )\!+\!2 \varphi \!-\!k\big],\\
	 0\!&\!=\!&\!(k-2 \varphi)\big(3 k^2-4 k \varphi -4 \varphi ^2+16 \varphi '\big);\\
	\text{Case B:~}  0\!&\!=\!&\!(\beta  k \tanh (k r)-2 \varphi )\big[\beta  k^2 \text{sech}^2(k r) (-3 \beta \nonumber\\
	&&+3 \beta  \cosh (2 k r)-16)-8 \beta  k \varphi  \tanh (k r)\nonumber\\
	&&-8 \varphi ^2+32 \varphi '\big],\\
	 0\!&\!=\!&\!\beta  k \tanh (k r)-2 \varphi ;\\
	\text{Case C:~} 0\!&\!=\!&\!F_{1}(\varphi,\varphi'),\\
	0\!&\!=\!&\!F_{2}(\varphi).
\end{eqnarray}
\end{subequations}
Here some redundant components of the field equations are not shown. Note that the functions $F_{1}$ and $F_{2}$ are complicated, so we are not going to show them here. On the other hand, the scalar field $\varphi$ follows its own field equation~\eqref{r4DEGBsfq2}. With the three sets of solutions obtained in this section, a branch of the corresponding scalar field solution is then given as follows:
\begin{subequations}
	\begin{align}
		&\text{Case A:}& \varphi=&-k/2,\\
		&\text{Case B:}& \varphi=&-\beta  k \tanh (k r)/2,\\
		&\text{Case C:}& \varphi=&-\sigma(r+\sqrt{2\alpha}\,\sigma/2)^{-1}/2.
	\end{align}
\end{subequations}
By substituting these scalar solutions into the gravitational field equations~\eqref{r4EGBgfeq2} respectively, it is true that they are satisfied automatically. Thus the cosmic string solutions obtained in the novel 4D EGB theory can also solve the regularized 4D EGB theory as we expected. In other words, the on-shell condition of the regularized 4D EGB theory is satisfied with our solutions which are obtained in the novel 4D EGB theory. It then supports that the two 4D EGB theories are equivalent to each other in the static cylindrically symmetric spacetime~\eqref{cc1}.

\section{Conclusion}\label{sec4}

The recent studies on the novel 4D EGB theory showed that the GB term could contribute to the Einstein's field equations by introducing a redefinition $\alpha\rightarrow\alpha/(D-4)$ and taking the limit $D\to4$. It brings abundant novel features to many phenomena including but not limited to gravitational waves, black holes, and wormholes. However, the theory does not admit an intrinsically four-dimensional definition in general, and may only make sense in some selected highly-symmetric spacetimes, such as the static spherically symmetric spacetime and the FLRW spacetime ~\cite{Gurses1,Fernandes1,Hennigar1}. Some regularization produces were then proposed to address these issues~\cite{Fernandes2,Lu1,Hennigar1,Kobayashi1,Bonifacio1}. The on-shell action of the resultant regularized 4D EGB theory is expected to have the same form as the original theory. Thus these two theories maybe equivalent to each other, while the regularized version could bypass the above issues. However, such equivalence is symmetry-dependent. The equivalence was already proved in the static spherically symmetric spacetime~\cite{Lu1,Yang2}, while it was found to be broken down in the Taub-NUT metric and the rotating black hole metrics~\cite{Hennigar1}. A main topic is then to test whether such equivalence holds in the other spacetime beyond spherical symmetry.

In this paper, we investigated the equivalence of these two theories in a static  spacetime with lower symmetry~\eqref{cc1}  compared with the spherically symmetric spacetime. We started from a higher-dimensional version of the EGB theory and derived the higher-dimensional field equations. We then redefined the GB coupling constant $\alpha\rightarrow\alpha/(D-4)$ and took the limit $D\rightarrow4$. Well-defined 4D field equations were obtained through this procedure. We further proved that, by eliminating the additional conformal scalar field with a branch of solution \eqref{onshell1}, the field equations of the two theories are exactly the same. So the static cylindrically symmetric metric~\eqref{cc1} can solve both the two theories. It supports that the two theories are equivalent to each other in such spacetime.

As some specific examples, we then investigated $D$-dimensional cosmic strings in the novel EGB theory. Three sets of cosmic string solutions were found. The 4D cosmic string solutions were obtained by setting $D\to4$, and their mass densities were also discussed. Therein, we found that when the cosmic string gives an asymptotically Minkowski spacetime, the nonvanishing GB term could remove the singularity located at the $z$ axis. We also showed that the GB term contributes nontrivial effect to the 4D cosmic string mass density. It is then expected that the GB term will affect the energy scale of the associate phase transition. It is worth to note that, in each set of our solution, the spacetime dimension $D$ is not fixed. Since in a higher-dimensional spacetime with $D\geq6$, the topological deflect (cosmic string) corresponds to a $d=(D-2)$-dimensional slice $\mathcal{M}_{d}$ with our Universe being a flat brane $\mathcal{M}_{4}$ embedded inside it,  our solutions are also useful to construct the braneworld model. Last, we used these solutions as some particular examples to support the equivalence of the novel 4D EGB theory and its regularized version in the static cylindrically symmetric spacetime~\eqref{cc1}.

Note that, throughout this paper, the metric~\eqref{metric1} corresponds to a specific case of the static cylindrically symmetric spacetime. Though the equivalence of the two theories was proved under the $D\rightarrow4$ version of such spacetime, one is still required to test it in a more general case. Besides, our proof is only valid at the  level of the background spacetime. No perturbations are involved in this paper. However, we know that the equivalence of these two theories is symmetry-dependent and that the perturbed spacetime is much less symmetric. Thus, if such equivalence breaks down in the general cylindrically symmetric spacetime, it is then expected to fail at the perturbation level even in our specific static cylindrically symmetric spacetime.

\section*{Acknowledgements}

We thank Si-Jiang Yang, Hao Yu, and Fei Qu for useful discussions. This work was supported by the National Natural Science Foundation of China (Grants No.~11875151, No.~11675064, and No.~11522541), the Strategic Priority Research Program on Space Science, the Chinese Academy of Sciences (Grant No.~XDA15020701), and the Fundamental Research Funds for the Central Universities (Grants No.~lzujbky-2020-it04 and No.~lzujbky-2019-ct06).


\begin{thebibliography}{1}


\bibitem{Tomozawa2011}
Y.~Tomozawa,
{\em Quantum corrections to gravity},
arXiv:1107.1424 [gr-qc].

\bibitem{Cognola2013}
G.~Cognola, R.~Myrzakulov, L.~Sebastiani, and S.~Zerbini,
{\em Einstein gravity with Gauss-Bonnet entropic corrections},
Phys.\ Rev.\ D\ \textbf{88}, 024006 (2013)
[arXiv:1304.1878 [gr-qc]].

\bibitem{Glavan1}
D.~Glavan and C.~Lin,
{\em Einstein-Gauss-Bonnet gravity in 4-dimensional space-time},
Phys.\ Rev.\ Lett.\ \textbf{124}, 081301 (2020)
[arXiv:1905.03601 [gr-qc]].




\bibitem{Konoplya1}
R.A.~Konoplya and A.F.~Zinhailo,
{\em Quasinormal modes, stability and shadows of a black hole in the novel 4D Einstein-Gauss-Bonnet gravity},
Eur.\ Phys.\ J.\ C {\bf 80}, 1049 (2020)
[arXiv:2003.01188 [gr-qc]].

\bibitem{Guo1}
M.~Guo and P.C.~Li,
{\em The innermost stable circular orbit and shadow in the novel $4D$ Einstein-Gauss-Bonnet gravity},
Eur.\ Phys.\ J.\ C\ \textbf{80}, 588 (2020)
[arXiv:2003.02523 [gr-qc]].

\bibitem{Fernandes1}
P.G.S.~Fernandes,
{\em Charged Black Holes in AdS Spaces in $4D$ Einstein Gauss-Bonnet Gravity},
Phys.\ Lett.\ B\ \textbf{805}, 135468 (2020)
[arXiv:2003.05491 [gr-qc]].

\bibitem{Wei1}
S.-W.~Wei and Y.-X.~Liu,
{\em Testing the nature of Gauss-Bonnet gravity by four-dimensional rotating black hole shadow},
arXiv:2003.07769 [gr-qc].

\bibitem{Zhang1}
Y.-P.~Zhang, S.-W.~Wei and Y.-X.~Liu,
{\em Spinning test particle in four-dimensional Einstein-Gauss-Bonnet Black Hole},
Universe {\bf 6}, 103 (2020),
[arXiv:2003.10960 [gr-qc]].

\bibitem{Lu1}
H.~Lu and Y.~Pang,
{\em Horndeski Gravity as $D\rightarrow4$ Limit of Gauss-Bonnet},
Phys.\ Lett.\ B {\bf 809}, 135717 (2020)
[arXiv:2003.11552 [gr-qc]].

\bibitem{Odintsov1}
S.D.~Odintsov, V.K.~Oikonomou, and F.P.~Fronimos,
{\em Rectifying Einstein-Gauss-Bonnet Inflation in View of GW170817},
Nucl.\ Phys.\ B {\bf 958}, 115135 (2020)
[arXiv:2003.13724 [gr-qc]].


\bibitem{Odintsov2}
S.D.~Odintsov and V.K.~Oikonomou,
{\em Swampland Implications of GW170817-compatible Einstein-Gauss-Bonnet Gravity},
Phys.\ Lett.\ B \textbf{805}, 135437 (2020),
[arXiv:2004.00479 [gr-qc]].

\bibitem{Heydari-Fard1}
M.~Heydari-Fard, M.~Heydari-Fard, and H.R.~Sepangi,
{\em Bending of light in novel 4$D$ Gauss-Bonnet-de Sitter black holes by Rindler-Ishak method},
arXiv:2004.02140 [gr-qc].

\bibitem{Yang1}
S.-J.~Yang, J.-J.~Wan, J.~Chen, J.~Yang, and Y.-Q.~Wang,
{\em Weak cosmic censorship conjecture for the novel $4D$ charged Einstein-Gauss-Bonnet black hole with test scalar field and particle},
Eur.\ Phys.\ J.\ C {\bf 80}, 937 (2020)
[arXiv:2004.07934 [gr-qc]].

\bibitem{Mahapatra1}
S.~Mahapatra,
{\em A note on the total action of $4D$ Gauss-Bonnet theory},
Eur.\ Phys.\ J.\ C {\bf 80}, 992 (2020)
[arXiv:2004.09214 [gr-qc]].

\bibitem{Hennigar1}
R.A.~Hennigar, D.~Kubiznak, R.B.~Mann, and C.~Pollack,
{\em On Taking the $D\to 4$ limit of Gauss-Bonnet Gravity: Theory and Solutions},
JHEP {\bf 2007}, 027 (2020)
[arXiv:2004.09472 [gr-qc]].




\bibitem{Jusufi2}
K.~Jusufi, A.~Banerjee, and S.G.~Ghosh,
{\em Wormholes in the novel 4D Einstein-Gauss-Bonnet Gravity},
Eur.\ Phys.\ J.\ C {\bf 80}, 698 (2020) 
[arXiv:2004.10750 [gr-qc]].


\bibitem{Hennigar2}
R.A.~Hennigar, D.~Kubiznak, R.B.~Mann, and C.~Pollack,
{\em Lower-dimensional Gauss-Bonnet Gravity and BTZ Black Holes},
Phys.\ Lett.\ B {\bf 808}, 135657 (2020)
[arXiv:2004.12995 [gr-qc]].


\bibitem{Oikonomou1}
V.~Oikonomou and F.~Fronimos,
{\em Reviving non-Minimal Horndeski-like Theories after GW170817: Kinetic Coupling Corrected Einstein-Gauss-Bonnet Inflation},
arXiv:2006.05512 [gr-qc].

\bibitem{Yang2}
K.~Yang, B.-M.~Gu, S.-W.~Wei, and Y.-X.~Liu,
{\em Born-Infeld Black Holes in novel 4D Einstein-Gauss-Bonnet gravity},
Eur.\ Phys.\ J.\ C {\bf 80}, 662 (2020) 
[arXiv:2004.14468 [gr-qc]].

\bibitem{Hennigar3}
R.A.~Hennigar, D.~Kubiznak, and R.B.~Mann,
{\em Rotating Gauss-Bonnet BTZ Black Holes},
arXiv:2005.13732 [gr-qc].

\bibitem{Easson1}
D.A.~Easson, T.~Manton, and A.~Svesko,
{\em $D\to4$ Einstein-Gauss-Bonnet Gravity and Beyond},
JCAP {\bf 2010}, 026 (2020)
[arXiv:2005.12292 [hep-th]].
  


\bibitem{Gurses1}
M.~Gurses, T.C.~Sisman, and B.~Tekin,
{\em Is there a novel Einstein-Gauss-Bonnet theory in four dimensions?},
Eur.\ Phys.\ J.\ C {\bf 80}, 647 (2020)
[arXiv:2004.03390 [gr-qc]].

\bibitem{Tian2020}
S.~Tian and Z.-H. Zhu,
{\em Non-full equivalence of the four-dimensional Einstein-Gauss-Bonnet gravity and Horndeksi gravity for Bianchi type I metric},
arXiv:2004.09954 [gr-qc].

\bibitem{Fernandes2}
P.G.S.~Fernandes, P.~Carrilho, T.~Clifton, and D.J.~Mulryne,
{\em Derivation of Regularized Field Equations for the Einstein-Gauss-Bonnet Theory in Four Dimensions},
Phys.~Rev.~D \textbf{102}, 024025 (2020)
[arXiv:2004.08362 [gr-qc]].

\bibitem{Kobayashi1}
T.~Kobayashi,
{\em Effective scalar-tensor description of regularized Lovelock gravity in four dimensions},
JCAP \textbf{07}, 013 (2020)
[arXiv:2003.12771 [gr-qc]].

\bibitem{Bonifacio1}
J.~Bonifacio, K.~Hinterbichler, and L.A.~Johnson,
{\em Amplitudes and 4D Gauss-Bonnet Theory},
Phys.~Rev.~D \textbf{102}, 024029 (2020)
[arXiv:2004.10716 [hep-th]].

\bibitem{Aoki1}
K.~Aoki, M.A.~Gorji, and S.~Mukohyama,
{\em A consistent theory of $D\rightarrow 4$ Einstein-Gauss-Bonnet gravity},
Phys.\ Lett.\ B {\bf 810}, 135843 (2020)
[arXiv:2005.03859 [gr-qc]].

\bibitem{Ai1}
W.Y.~Ai,
{\em A note on the novel 4D Einstein-Gauss-Bonnet gravity},
Commun.\ Theor.\ Phys.  {\bf 72}, 095402 (2020),
[arXiv:2004.02858 [gr-qc]].

\bibitem{Kirzhnits1}
D.A.~Kirzhnits and A.D.~Linde,
{\em On the Vacuum Stability Problem in Quantum Electrodynamics},
Phys.~Lett.~B \textbf{73}, 323 (1978).

\bibitem{Vilenkin1}
A.~Vilenkin and E.P.S.~Shellard,
{\em Cosmic strings and other topological defects}, Cambridge Monographs on
Mathematical Physics, Cambridge, 2001.


\bibitem{Kaluza1}
T.~Kaluza,
{\em Zum Unit\"{a}tsproblem der Physik},
Sitzungsber.~Preuss.~Akad.~Wiss.~Berlin (Math.~Phys.) \textbf{1921}, 966 (1921) [Int.~J.~Mod.~Phys.~D \textbf{27}, 1870001 (2018)]
[arXiv:1803.08616 [physics.hist-ph]].

\bibitem{Klein1}
O.~Klein,
{\em Quantum Theory and Five-Dimensional Theory of Relativity} (In German and English),
Z.~Phys. \textbf{37}, 895 (1926) [Surveys High Energ.~Phys. \textbf{5}, 241 (1986)].

\bibitem{Klein2}
O.~Klein,
{\em The Atomicity of Electricity as a Quantum Theory Law},
Nature \textbf{118}, 516 (1926).

\bibitem{Dvali1}
G.R.~Dvali, G.~Gabadadze, and M.~Porrati,
{\em 4-D gravity on a brane in 5-D Minkowski space},
Phys.~Lett.~B \textbf{485}, 208 (2000)
[hep-th/0005016].

\bibitem{Deffayet1}
C.~Deffayet and K.~Menou,
{\em Probing Gravity with Spacetime Sirens},
Astrophys.~J. \textbf{668}, L143 (2007)
[arXiv:0709.0003 [astro-ph]].

\bibitem{Randall1}
L.~Randall and R.~Sundrum,
{\em A Large mass hierarchy from a small extra dimension},
Phys.\ Rev.\ Lett. \textbf{83}, 3370 (1999)
[hep-ph/9905221].

\bibitem{Randall2}
L.~Randall and R.~Sundrum,
{\em An Alternative to compactification},
Phys.\ Rev.\ Lett. \textbf{83}, 4690 (1999)
[hep-th/9906064].

\bibitem{Olasagasti1}
I.~Olasagasti and A.~Vilenkin,
{\em Gravity of higher dimensional global defects},
Phys.~Rev.~D \textbf{62}, 044014 (2000)
[hep-th/0003300].

\bibitem{Gherghetta1}
T.~Gherghetta and M.E.~Shaposhnikov,
{\em Localizing gravity on a string-like defect in six-dimensions},
Phys.\ Rev.\ Lett. \textbf{85}, 240 (2000)
[hep-th/0004014].

\bibitem{Oda1}
I.~Oda,
{\em Localization of matters on a string-like defect},
Phys.\ Lett.\ B \textbf{496}, 113 (2000)
[hep-th/0006203].

\bibitem{Dzhunushaliev1}
V.~Dzhunushaliev, V.~Folomeev, and M.~Minamitsuji,
{\em Thick brane solutions},
Rept.\ Prog.\ Phys. \textbf{73}, 066901 (2010)
[arXiv:0904.1775 [gr-qc]].

\bibitem{Dadhich1}
N.~Dadhich, S.G.~Ghosh, and S.~Jhingan,
{\em The Lovelock gravity in the critical spacetime dimension},
Phys.\ Lett.\ B \textbf{711}, 196-198 (2012)
[arXiv:1202.4575 [gr-qc]].


\bibitem{Hogan1}
C.J.~Hogan and M.J.~Rees,
{\em Gravitational interactions of cosmic strings},
Nature \textbf{311}, 109 (1984).

\bibitem{Vachaspati1}
T.~Vachaspati and A.~Vilenkin,
{\em Gravitational Radiation from Cosmic Strings},
Phys.\ Rev.\ D \textbf{31}, 3052 (1985).

\bibitem{Bennett1}
D.P.~Bennett and F.R.~Bouchet,
{\em Evidence for a Scaling Solution in Cosmic String Evolution},
Phys.\ Rev.\ Lett. \textbf{60}, 257 (1988).

\bibitem{Caldwell1}
R.R.~Caldwell and B.~Allen,
{\em Cosmological constraints on cosmic string gravitational radiation},
Phys.\ Rev.\ D \textbf{45}, 3447 (1992).

\bibitem{Kaiser1}
N.~Kaiser and A.~Stebbins,
{\em Microwave Anisotropy Due to Cosmic Strings},
Nature \textbf{310}, 391 (1984).

\bibitem{Vachaspati2}
T.~Vachaspati,
{\em Cosmic Strings and the Large-Scale Structure of the Universe},
Phys.\ Rev.\ Lett. \textbf{57}, 1655 (1986).

\bibitem{Danos1}
R.J.~Danos, R.H.~Brandenberger, and G.~Holder,
{\em A Signature of Cosmic Strings Wakes in the CMB Polarization},
Phys.\ Rev.\ D \textbf{82}, 023513 (2010)
[arXiv:1003.0905 [astro-ph.CO]].

\bibitem{Brandenberger1}
R.H.~Brandenberger, R.J.~Danos, O.F.~Hernandez, and G.P.~Holder,
{\em The $21$\,cm Signature of Cosmic String Wakes},
JCAP \textbf{1012}, 028 (2010)
[arXiv:1006.2514 [astro-ph.CO]].

\bibitem{Vilenkin2}
A.~Vilenkin,
{\em Cosmic strings as gravitational lenses},
Astrophys.\ J. \textbf{282}, L51 (1984).

\bibitem{Jusufi1}
K.~Jusufi and A.~Ovgün,
{\em Effect of the cosmological constant on the deflection angle by a rotating cosmic string},
Phys.\ Rev.\ D \textbf{97}, 064030 (2018)
[arXiv:1712.01771 [gr-qc]].


\bibitem{Ovgun1}
A.~Övgün,
{\em Weak field deflection angle by regular black holes with cosmic strings using the Gauss-Bonnet theorem},
Phys.\ Rev.\ D \textbf{99}, 104075 (2019)
[arXiv:1902.04411 [gr-qc]].

\bibitem{Yamauchi1}
D.~Yamauchi, T.~Namikawa, and A.~Taruya,
{\em Weak lensing generated by vector perturbations and detectability of cosmic strings},
JCAP \textbf{1210}, 030 (2012)
[arXiv:1205.2139 [astro-ph.CO]].


\bibitem{Hammond1}
D.K.~Hammond, Y.~Wiaux, and P.~Vandergheynst,
{\em Wavelet domain Bayesian denoising of string signal in the cosmic microwave background},
Mon.\ Not.\ Roy.\ Astron.\ Soc. \textbf{398}, 1317 (2009)
[arXiv:0811.1267 [astro-ph]].

\bibitem{Blanco-Pillado1}
J.J.~Blanco-Pillado, K.D.~Olum, and X.~Siemens,
{\em New limits on cosmic strings from gravitational wave observation},
Phys.\ Lett.\ B \textbf{778}, 392 (2018)
[arXiv:1709.02434 [astro-ph.CO]].

\bibitem{Ringeval1}
C.~Ringeval and T.~Suyama,
{\em Stochastic gravitational waves from cosmic string loops in scaling},
JCAP \textbf{1712}, 027 (2017)
[arXiv:1709.03845 [astro-ph.CO]].

\bibitem{Sadr1}
A.V.~Sadr, S.M.~S.~Movahed, M.~Farhang, C.~Ringeval, and F.R.~Bouchet,
{\em A Multiscale pipeline for the search of string-induced CMB anisotropies},
Mon.\ Not.\ Roy.\ Astron.\ Soc. \textbf{475}, 1010 (2018)
[arXiv:1710.00173 [astro-ph.CO]].


\bibitem{Ade1}
P.A.R.~Ade {\em et al.} [Planck Collaboration],
{\em Planck 2013 results. XXV. Searches for cosmic strings and other topological defects},
Astron.\ Astrophys. \textbf{571}, A25 (2014)
[arXiv:1303.5085 [astro-ph.CO]].

\bibitem{Abbott1}
B.P.~Abbott {\em et al.} [LIGO Scientific and Virgo Collaborations],
{\em Constraints on cosmic strings using data from the first Advanced LIGO observing run},
Phys.\ Rev.\ D \textbf{97}, 102002 (2018)
[arXiv:1712.01168 [gr-qc]].

\bibitem{Abbott2}
B.P.~Abbott {\em et al.} [LIGO Scientific and Virgo Collaborations],
{\em Search for the isotropic stochastic background using data from Advanced LIGO’s second observing run},
Phys.\ Rev.\ D \textbf{100}, 061101 (2019)
[arXiv:1903.02886 [gr-qc]].

\bibitem{Gregory1}
R.~Gregory,
{\em Nonsingular global string compactifications},
Phys.\ Rev.\ Lett. \textbf{84}, 2564 (2000)
[hep-th/9911015].

\bibitem{Momeni1}
D.~Momeni and H.~Gholizade,
{\em A note on constant curvature solutions in cylindrically symmetric metric $f(R)$ Gravity},
Int.\ J.\ Mod.\ Phys.\ D \textbf{18}, 1719 (2009)
[arXiv:0903.0067 [gr-qc]].

\bibitem{Sharif1}
M.~Sharif and S.~Arif,
{\em Static cylindrically symmetric interior solutions in $f(R)$ gravity},
Mod.\ Phys.\ Lett.\ A \textbf{27}, 1250138 (2012)
[arXiv:1302.1191 [gr-qc]].








\end{thebibliography}
\end{document}